\documentclass[epsfig]{article}
\usepackage{amssymb}
\usepackage{graphicx}
\usepackage{amsmath}

\input epsf.sty
\newtheorem{theorem}{Theorem}
\newtheorem{acknowledgement}[theorem]{Acknowledgement}

\input{tcilatex}
\makeatletter \@addtoreset{equation}{section}
\renewcommand{\theequation}{\thesection.\arabic{equation}}

\begin{document}

\title{\rightline{\mbox {\normalsize
{Lab/UFR-HEP0401/GNPHE/0401}}} \textbf{Hyperbolic Invariance}}
\author{Malika Ait Ben Haddou$^{1,2,3}$\thanks{%
aitbenha@fsmek.ac.ma} and El Hassan Saidi$^{2,3}$\thanks{%
h-saidi@fsr.ac.ma} \\
{\small 1 D\'{e}partement de Math\'{e}matique \& Informatique, Facult\'{e}
des Sciences, Meknes, Morocco.}\\
{\small 2}. {\small Lab/UFR-Physique des Hautes Energies, Facult\'{e} des
Sciences de Rabat, Morocco.}\\
{\small 3-Groupement National de Physique des Hautes Energies, GNPHE; Siege
focal, Rabat, Morocco.}}
\maketitle

\begin{abstract}
Motivated by the study of duality cascades in supersymmetric quiver gauge
theories beyond affine models, we develop in this paper the analysis of a class of simply laced hyperbolic Lie
algebras. These are specific generalizations of affine ADE symmetries which
form a particular subclass of the so-called Indefinite Lie algebras.
Because of indefinite signature of their bilinear form, we show that these
infinite dimensional invariances have very special features and admit a remarkable
link type IIB background with non zero axion. We also show that
hyperbolic root system $\Delta _{hyp}$ has a $\mathbb{Z}_{2}\mathbb{\times Z}%
_{3}$ gradation containing two specific and isomorphic proper subsets of
affine Kac-Moody root systems baptized as $\Delta _{affine}^{\delta }$ and \
$\Delta _{affine}^{\gamma }$. We give an explicit form of the
commutation relations for hyperbolic ADE algebras and analyze their Weyl groups W$%
_{hyp}$. Comments regarding links with Seiberg like dualities and RG
cascades are made.

\end{abstract}

 \thispagestyle{empty} \setcounter{page}{1}

 \newpage
  \tableofcontents
  \newpage

\section{Introduction}

During last few decades, finite dimensional Lie algebras and affine
Kac-Moody (KM) extensions as well as their representations have been
intensively used in establishing strong results in quantum field theory.
Recently, root systems and Weyl symmetries of these invariances together
with their algebraic geometry analog have been shown to be behind the
developments made in the study of supersymmetric quiver gauge theories
viewed as type II string low energy limit. In fact this remarkable link
between Lie algebra roots and quantum fields is not a new thing since it
goes back to the early days of discovery of gauge theories and turns out to
encode most of their basic features. Roots are explicitly manifest in
several quantum field theoretic models such in the study of exactly solvable
hamiltonian systems, integrable Toda field models $\cite{a1}$-$\cite{a11}$,
conformal field theories (CFTs) in low and higher dimensions $\cite{b12}$-$%
\cite{b19}$ and string theory.

In four dimensional supersymmetric QFTs, the algebraic features of roots of
simply laced ADE Lie algebras play an important role in the study of QFT$%
_{4} $ class embedded in type II string compactifications on local
Calabi-Yau (CY) threefolds. More precisely, they are used in the geometric
engineering of supersymmetric quiver gauge theories $\cite{c1}$-$\cite{c10}$
and in the study of their D-brane realizations. These supersymmetric quiver
QFT$_{4}$s, which appear as specific QFT limits of type II string on K3
fibered CY manifolds with ordinary and affine ADE geometries $\cite{c11}$,
are nowadays subject of great interest in connection with large N field and
string dualities$\cite{c12}$-$\cite{c17}$.

The principal aim of the present study is to extend results on ordinary ADE
and their affine KM structures, as used in $4D$ $\mathcal{N}=2$ and $%
\mathcal{N}=1$ quiver gauge theories, to the case of hyperbolic
generalization of ADE Lie algebras. This generalization is the next leading
extension of ordinary ADE Lie algebras which, surprisingly enough, haven't
been sufficiently explored in literature. The leading extension of ordinary
ADE is naturally the affine Kac-Moody ADE algebras. Their Lie algebra and
algebraic geometry properties as well as the role they play in
supersymmetric quiver QFT$_{4}$s are now quite well established.

As we know, there are only partial algebraic results on the indefinite
sector of Lie algebras, for instance a classification \`{a} la Cartan of
these algebras is still missing. So to address the objectives of this study,
we first have to complete results on hyperbolic Lie algebras and their
algebraic geometry analog. Once armed with these mathematical results, we
can then consider the physical application. To avoid a presentation with lot
of technicalities mixing physics and diverse mathematical methods, we have
judged instructive to divide this analysis into two parts. The first part,
to be developed in this paper, deals with the study of roots and Weyl
symmetries in hyperbolic algebras. In part II $\cite{brs}$, we consider the
geometric engineering of 4D $\mathcal{N}=2$ and $\mathcal{N}=1$\ quiver QFTs
based on these hyperbolic ADE Lie algebras as well as their D brane
realizations. The results obtained in present article will be also used to
study the QFT duals of hyperbolic quiver gauge theories as well as the
analog of RG cascades of affine models.

The organization of this paper is as follows. In section 2, we give details
on motivations in our interest into hyperbolic quiver gauge theories. In
section 3, we review general aspects of Indefinite Lie algebras; in
particular their special hyperbolic subset. Main interest is focused on
simply laced hyperbolic ADE symmetries seen the role they play in the
construction of hyperbolic quiver QFTs. In section 4, we study the root
systems $\Delta _{hyp}$ of these hyperbolic ADE Lie algebras and derive the
explicit contents of $\Delta _{hyp}$ as well as their closed proper subsets.
In section 5, we summarize our results on $\Delta _{hyp}$\ into a theorem
giving the full set of root contents of hyperbolic ADE algebras and a
corollary on the way they may be used. In this regards it is interesting to
anticipate on a particular result in hyperbolic ADE extension which turns
out to be very helpful for the building of the hyperbolic structure. As we
will show, hyperbolic ADE have two special isomorphic affine ADE subalgebras
denoted as \textbf{g}$_{affine}^{\delta }$ and \textbf{g}$_{affine}^{\gamma
} $ and most of properties of hyperbolic symmetry may be viewed as kinds of
interpolations between corresponding properties into these two specific
affine subalgebras. In section 6, we derive the explicit form of the
commutation relations for hyperbolic ADE Lie algebras using first the
interpolation method between the commutation relations of \textbf{g}$%
_{affine}^{\delta }$ and \textbf{g}$_{affine}^{\gamma }$ and second by
introducing a covariant construction based on the use of the bi-linear form
of the hyperbolic ADE algebras. In this section, we also give the necessary
and sufficient conditions for the unitary highest weight representations of
hyperbolic ADE algebras. In section 7, we construct Weyl groups W$_{hyp}$ of
these hyperbolic algebras using interpolation scenario between the Weyl
sub-groups associated with \textbf{g}$_{affine}^{\delta }$ and \textbf{g}$%
_{affine}^{\gamma }$. Last section is devoted to conclusion and comments.

\section{Motivations}

In this section, we make two specific comments to motivate our interest into
hyperbolic ADE extensions of supersymmetric quiver gauge theories. The first
one deals with the algebraic geometry interpretation of ordinary and affine
quiver gauge theories and the second with their large N dualities. Then, we
address the question of building hyperbolic quiver gauge model. More details
on this issue as well as their link with type IIB background with non zero
axion can be found in second paper \cite{brs}.

\subsection{Roots in quiver gauge models}

To begin recall that 4D supersymmetric ADE quiver gauge theories are QFT$%
_{4} $ limits of type II strings on CY threefolds with ADE geometries and
are remarkably engineered on ADE Dynkin diagrams. Nodes of the Dynkin graphs
encode gauge and adjoint matter multiplets and links between the nodes
engineer the various kinds of bi-fundamental matters involved in the
supersymmetric $\prod_{i}U\left( N_{i}\right) $ quiver gauge theory.
Moreover, in these 4D quiver QFTs, roots $\alpha $ ($\alpha =\pm \sum
k_{i}\alpha _{i}$, $k_{i}\in \mathbb{Z}_{+}$) of ADE Lie algebras generated
by the $\alpha _{i}$ simple ones, generally realized in $\mathbb{R}^{n}=\sum
\mathbb{R}e_{i}$ ($e_{i}e_{j}=\delta _{ij}$) like%
\begin{equation}
\alpha _{i}=e_{i}-e_{i+1},\qquad i=1,...,
\end{equation}%
as in case of $su\left( n\right) $ Lie algebras, have an algebraic geometry
interpretation. They are in one to one correspondence with the holomorphic
volumes
\begin{equation}
\upsilon _{i}=\int_{CP_{i}^{1}}\Omega ^{\left( 2,0\right) },
\end{equation}
of\ the homological two-cycles CP$_{i}^{1}$ involved in the resolution of
ADE singularities $\cite{d1}$,
\begin{equation}
\upsilon _{i}=t_{i}-t_{i+1},\qquad i=1,....
\end{equation}%
In this relation the $t_{i}$s are complex moduli and the $\upsilon _{i}$s
describe deformations of local ADE geometry and have a nice interpretation
in 4D $\mathcal{N}=2$ supersymmetric quiver $\Pi _{i=1}^{r}U\left(
N_{i}\right) $ gauge theories with adjoint matter superfields $\left\{ \Phi
_{i}\quad i=1,...,r\right\} $. There, the $\upsilon _{i}$\ moduli appear as
FI like couplings generating the following 4D $\mathcal{N}=1$ linear chiral
superspace potential deformation $\mathrm{\delta W}$,
\begin{equation}
\mathrm{\delta W}=\sum_{i=1}^{r}\upsilon _{i}\int d^{4}xd^{2}\theta Tr\left(
\Phi _{i}\right) \text{.}  \label{dw}
\end{equation}%
This special superpotential deformation preserves $\mathcal{N}=2$
supersymmetry and its non linear $\mathcal{N}=1$ extension is at the basis
of the field theoretic representation of the geometric transition $O\left(
-1\right) \times O\left( -1\right) \times CP^{1}\rightarrow T^{\ast }S^{3}$
of the conifold. Eq(\ref{dw}) is also behind the field theoretic analysis of
large $N$\ field dualities and in the derivation of exact results in $%
\mathcal{N}=1$ supersymmetric gauge theories $\cite{d1,d2}$-$\cite{c12}$.

The second comment we want to make\ is about quantum field interpretation of
automorphism symmetry of root system of ADE Lie algebras. Like for roots $%
\alpha $, Weyl groups $W_{ADE}$ of Lie algebras play also a crucial role in
understanding part of quantum field theoretic dualities; in particular
Seiberg like dualities and RG cascades of affine models $\cite{c8}$. At low
energies below string scale where the dynamics of matter and gauge fields is
governed by supersymmetric Yang Mills model, one disposes of sets\ of dual
ADE quiver gauge theories with a remarkable subclass whose duality
symmetries act on previous $\upsilon _{i}$ as
\begin{equation}
\upsilon _{i}\rightarrow \upsilon _{i}^{\prime }=A_{ij}\upsilon _{j},
\label{v}
\end{equation}%
These duality symmetries were shown to be isomorphic to the usual Weyl group
transformations of ADE root system $\cite{c8}$. By help of correspondence (%
\ref{v}), the $A_{ij}$\ matrix in above relation is isomorphic to the
bi-linear product\ $\delta _{ij}-\alpha _{i}\alpha _{j}$ that appears in
Weyl reflections $\alpha _{i}^{\prime }=\alpha _{i}-2\frac{\alpha _{i}\alpha
_{j}}{\alpha _{j}^{2}}\alpha _{j}$.

In addition to the two above links, there are other basic ties between 4D
super quiver QFTs and ADE algebra. For instance ADE root systems and their
Weyl symmetries are also used in brane realization of the quiver gauge
theories living in the world volume of parallel $N_{0}$ D3 branes and $N$ D5
ones partially wrapping CP$_{i}^{1}$ two-cycles of CY3 folds with a local
ADE geometry. There, $N_{0}$ D3 is roughly speaking associated with the
affine simple root $\alpha _{0}$ of affine KM ADE root system and wrapped $N$
D5s with remaining $\alpha _{i}$ simple ones. In this representation, field
theoretic scenarios such as higgsings correspond just to special properties
of the root system. Other basic relations between roots and their Weyl
automorphisms on one hand; and relevant QFT$_{4}$ moduli on the other hand
can be also written down. Supersymmetric Yang-Mills gauge couplings g$%
_{i}^{SYM}$ of the quiver gauge sub-group factors $U\left( N_{i}\right) $
and corresponding beta functions $b_{i}$ including supersymmetric affine ADE
conformal field models,
\begin{equation}
\frac{1}{g_{s}}=\sum_{i=0}^{r}\varepsilon _{i}g_{i}^{-2};\qquad
b_{D}=\sum_{i=0}^{r}\varepsilon _{i}b_{i},
\end{equation}%
with $\varepsilon _{i}$s the usual Dynkin weights, obey a similar law as
holomorphic volumes $\upsilon _{i}$ eq(\ref{v}). For details on this issue
as well as other areas of involvement of Weyl symmetries, we refer to $\cite%
{brs}$, see also $\cite{e1}$-$\cite{e3}$.

\subsection{Superfield action}

Despite that above quiver gauge field theories based on finite ADE and
affine ADE Lie algebras belong to different classes of QFT$_{4}$s ( affine
QFTs are CFTs while ordinary ones are generally not) and though they have
different physical interpretations and different D branes realization; they
do however share most of basic features. The point is that their
fundamentals are quite same and the observed physical disparities are
nothing but manifestations of the Lie algebraic deformations.
\begin{equation}
\text{\textit{affine ADE QFT}}_{4}\text{\textit{s} \ }\sim \text{\ \ \textit{%
ordinary ADE QFT}}_{4}\text{\textit{s} \ plus \ \textit{special deformations.%
}}  \label{e}
\end{equation}%
From this vision of things, one clearly see that basic properties shared by
these two classes of quiver QFT$_{4}$s are in fact just a part of a general
picture involving larger extensions of ADE symmetries. This behaviour should
be also valid for other extensions of ADE symmetries; in particular for
hyperbolic ADE symmetry we are interested in here. As such previous
correspondence extends naturally to,
\begin{equation}
\text{\textit{hyperbolic ADE QFT}}_{4}\text{\textit{s} \ }\sim \text{\ \
\textit{affine ADE QFT}}_{4}\text{\textit{s \ }plus\textit{\ \ appropriate
deformations},}
\end{equation}%
forming together with eq(\ref{e}) a sequence of three terms describing the
first two leading deformations of ordinary super quiver ADE QFT$_{4}$s.

Recall that from Lie algebraic point of view, affine ADE buildings may be
viewed as a leading extension of the corresponding ordinary ADE ones.
Starting from a rank $r$ Cartan matrix $K_{finite}$ of a given finite
dimensional ADE Dynkin diagram, this extension mainly consists to add a
special (affine) node to the ordinary Dynkin diagrams as,
\begin{equation}
K_{finite}\qquad \rightarrow \qquad K_{affine}=\left(
\begin{array}{cc}
2 & -1 \\
-1 & K_{finite}%
\end{array}
\right) ,
\end{equation}
where the $\left( -1\right) $ entry in the first row refers to the line $r$%
-vector $\left( -1,0,...,0\right) $ and the other to its transpose. Note
that for ADE cases, $K_{finite}$ matrix is symmetric and can be split in
general as $K_{finite}^{ij}=2\delta ^{ij}+2G_{f}^{\left( ij\right) }$ where
we have set $G_{f}^{\left( ij\right) }=\left( G_{f}^{ij}+G_{f}^{ji}\right)
/2 $ and $G_{f}^{ij}=G_{f}^{\left( ij\right) }+G_{f}^{\left[ ij\right] }$.
Similar decomposition can be also done for generalized affine ADE and
hyperbolic ADE Cartan matrices. We have then $K_{affine}^{ij}=2\delta
^{ij}+2G_{af}^{\left( ij\right) }$ and $K_{hyp}^{ij}=2\delta
^{ij}+2G_{hyp}^{\left( ij\right) }$.

In geometric engineering of four dimensional $\mathcal{N}=2$ ($\mathcal{N}=1$%
) supersymmetric ADE quiver gauge theories with adjoint $\Phi _{i}$ matters
and bi-fundamental $Q_{ij}$ ones, the above algebraic affine extension has a
superfield theory realization. It corresponds to deforming the ( massive)
ordinary ADE quiver theory described by the superfield action $\mathcal{S}%
_{finite}=\mathcal{S}_{finite}\left[ Q,V,\Phi \right] $,
\begin{eqnarray}
\mathcal{S}_{finite} &=&\int d^{4}xd^{4}\theta \left[ \sum_{i,j=1}^{r}Tr%
\left( Q_{ij}^{\ast }\left[ \exp \left( K_{finite}^{ij}V_{j}\right) \right]
Q_{ji}\right) \right] +\sum_{i=1}^{r}\zeta _{i}\int d^{4}xd^{4}\theta
Tr\left( V_{i}\right)  \notag \\
&&+\int d^{4}xd^{4}\theta \left[ \sum_{j=r+1}^{n}Tr\left( Q_{ij}^{\ast }%
\left[ \exp \left( q_{finite}^{ij}V_{j}\right) \right] Q_{ji}\right) \right]
+\sum_{i=r+1}^{n}\zeta _{i}\int d^{4}xd^{4}\theta Tr\left( V_{i}\right)
\notag \\
&&+\left( \int d^{4}xd^{2}\theta \left[ \sum_{i,j=1}^{r}G_{f}^{\left[ ij%
\right] }Tr\left( Q_{ij}\Phi _{j}Q_{ji}\right) +\sum_{i=1}^{r}Tr\mathrm{W}%
\left( \Phi _{i}\right) \right] +hc\right)  \label{act} \\
&&+\left( \int d^{4}xd^{2}\theta \left[ \sum_{i}Tr\left( \Phi _{i}^{\ast
}e^{K_{ii}^{q}adV_{i}}\Phi _{i}\right) \right] +\int d^{4}xd^{2}\theta \left[
\sum_{i=1}^{r}\frac{1}{g_{i}^{2}}Tr\left( W_{\alpha }^{i}W_{i}^{\alpha
}\right) \right] +hc\right) ,  \notag
\end{eqnarray}%
to a 4D $\mathcal{N}=2$ ($\mathcal{N}=1$)\ affine ADE quiver QFT involving
more gauge and matter superfields\footnote{%
The extra term in $\mathcal{S}_{finite}$ involving the rectangular matrix q$%
_{finite}^{ij}$ is required by type II string on CY3 with a local ADE
geometry. The explicit form of these q$^{ij}$s may be found in \cite{b12}-
\cite{b15}}. The extra terms with $\zeta _{i}$ couplings are the usual FI
terms and the $V_{i}$\ are the gauge superfields; $\left( n-r\right) $ of
them are auxiliary as they have no propagating dynamics, they are introduced
for technical reasons; in particular in order to ensure the CY condition in
string compactification. They are also needed in the study of the critical
behaviour of these supersymmetric quiver QFT$_{4}$s. The corresponding
superfield action $\mathcal{S}_{affine}$\ is obtained from previous one by
substituting $K_{finite}$, $G_{f}^{\left[ ij\right] }$ and $q_{finite}^{ij}$
by their affine $K_{affine}$, $G_{af}^{\left[ ij\right] }$ and $%
q_{affine}^{ij}$ analog. Since the action $\mathcal{S}_{affine}$ can be
usually put in the form,

\begin{equation}
\mathcal{S}_{affine}=\mathcal{S}_{finite}\text{ \ \ }+\text{ \ \
deformations \ \ ,}  \label{deff}
\end{equation}
one can easily identify the superfield operators that capture the
deformation from ordinary ADE quiver gauge theories to the affine ADE ones.
In this way, one can recover many known results on QFT deformations such for
instance supersymmetric Sine Gordon model and affine ADE Toda field theories
respectively obtained by deformations of supersymmetric Liouville theory and
ordinary ADE Toda models. A similar conclusion is also valid for affine ADE
CFTs and their underlying partners.

In D-brane realization of four dimensional QFT$_{4}$s living in the non
compact directions of D5 branes wrapping two cycles with the topology of
ordinary ADE Dynkin diagrams, the algebraic affine extension has also a D
brane interpretation. It corresponds to adding D3 branes to the existing
system of wrapped D5 ones as shown below$\cite{c8}$.
\begin{equation}
\cup _{i=1}^{r}\left\{ N_{i}D5/S_{i}^{2}\right\} \qquad \rightarrow \qquad
\cup _{i=1}^{r}\left\{ N_{i}D5/S_{i}^{2}\right\} \text{ \ \ }\cup \text{ \ \
}N_{0}D3.
\end{equation}%
Here the $S_{i}^{2}$ are the two cycles of the resolved ADE geometry; they
are in one to one with the simple roots $\alpha _{i}$ of ordinary ADE
algebras. Extra D3s fill the transverse space to the $S_{i}^{2}$s and have
much to do with the imaginary root $\delta =\sum_{i=0}^{r}\epsilon
_{i}\alpha _{i}$\ of affine ADE systems. The homological cycle associated
with $\delta $ is known to have no proper holomorphic volume.

\subsection{Beyond affine model}

From above quantum field analysis, one sees that the established results we
have for ordinary and affine quiver gauge models are in fact more general.
They may be extended for supersymmetric quiver gauge theories based on those
simply laced Lie algebras going beyond affine KM ones. These gauge theories
are also expected to follow as low energy type II string compactification on
a particular class of CY3s and also have specific D-brane realizations.

From Lie algebra view, one also see that infinite dimensional affine KM
generalization is not the unique possible extension of ordinary ADE Lie
algebras. The affine KM extension of ordinary Lie algebras is an interesting
generalization; but roughly speaking has nothing special except that it is
the leading one and has a zero eigenvalue. Actually there are many other
possible and remarkable generalizations although only few of them are under
control.

One of the objectives of this paper and its continuation \cite{brs} is to
work the complete picture by building consistent quantum field theoretical
model going beyond the affine ADE ones. To achieve this goal, it is
interesting to recall that many recent results regarding affine ADE field
models are approached through the algebraic properties of affine ADE root
systems $\Delta _{affine}$ and their algebraic geometry counterpart. Here
also we will follow this path to study hyperbolic extension of affine
models. To that purpose, we need first of all basic information on the
algebraic properties of extensions of affine ADE KM algebras. But
unfortunately and remarkably these are not fully available in literature.
Except for some examples $\cite{m1}$-$\cite{m01}$, to our knowledge the
explicit content of root system $\Delta _{hyp}$ of hyperbolic Lie algebras,
their corresponding commutation relations and associated Weyl symmetries
have not been yet completely identified. This is why we propose to first
work out explicitly the aforementioned structures for hyperbolic ADE Lie
algebras and then come back to the analysis of supersymmetric hyperbolic ADE
quiver gauge theories.

In what follows, we focus on the explicit derivation of useful tools on
hyperbolic ADE Lie algebras; in particular their root system, the
commutation relations, unitary conditions for highest weight representations
and Weyl symmetries. We show amongst others:\newline
(a) The root system $\Delta _{hyp}$ of hyperbolic extension of affine ADE
Lie algebras is given by,
\begin{equation}
\Delta _{hyp}\cup \left\{ 0\right\} =\left\{ n\gamma +m\delta +l\alpha
;\quad l^{2}-mn\leq 1\qquad l=0,\pm 1;\quad ;n,m\in \mathbb{Z\qquad }\alpha
\in \Delta _{finite}\right\} ,
\end{equation}%
where $\gamma $ and $\delta $ are two imaginary roots satisfying $\gamma
^{2}=\delta ^{2}=0$ and $\gamma \delta =-1$ and where $\Delta _{finite}$\ is
the root system of the underlying ordinary ADE Lie subalgebra.\newline
(b) The commutation\ relations defining the hyperbolic ADE Lie algebras read
as,%
\begin{eqnarray}
\left[ L,K\right] &=&\left[ K,\alpha H\right] =\left[ \alpha H,\beta H\right]
=0,\qquad \alpha ,\beta \in \Delta _{finite},  \notag \\
\left[ \alpha H,\mathrm{T}_{p,q}^{j\beta }\right] &=&j\left( \alpha \beta
\right) \text{ }\mathrm{T}_{p,q}^{\pm j\beta };\qquad \alpha ,\beta \in
\Delta _{finite}  \notag \\
\left[ K,\mathrm{T}_{p,q}^{j\beta }\right] &=&q\mathrm{T}_{p,q}^{j\beta
};\qquad \beta \in \Delta _{finite}  \notag \\
\left[ L,\mathrm{T}_{p,q}^{j\beta }\right] &=&p\mathrm{T}_{p,q}^{j\beta
};\qquad \beta \in \Delta _{finite}  \label{hypa} \\
\left[ \mathrm{T}_{m,n}^{l\alpha },\mathrm{T}_{p,q}^{j\beta }\right] &=&%
\frac{Y\left( l^{2}\alpha ^{2}-2mn-2\right) Y\left( j^{2}\alpha
^{2}-2pq-2\right) }{Y\left( \left( l\alpha +j\beta \right) ^{2}-2\left(
m+p\right) \left( n+q\right) -2\right) }\varepsilon _{l\alpha ,j\beta }%
\mathrm{T}_{m+p,n+q}^{\left( l\alpha +j\beta \right) };\qquad \alpha ,\beta
\in \Delta _{finite},  \notag \\
\left[ \mathrm{T}_{m,n}^{l\alpha },\mathrm{T}_{-m,-n}^{-l\alpha }\right] &=&%
\frac{2Y\left( l^{2}\alpha ^{2}-2mn-2\right) }{2mn-l^{2}\alpha ^{2}}\left(
nL+mK-l\alpha H\right) ,\qquad \alpha \in \Delta _{finite}.  \notag
\end{eqnarray}%
In these relations, the operators $L,K,\alpha H$ and $\mathrm{T}%
_{p,q}^{j\beta }$ are the generators of hyperbolic algebra; they will be
discussed in details in sub-section 6.3. The function $Y\left( x\right) =1$
if $x\leq 0$ and zero otherwise, is the Heveaside like distribution. Note
that setting $n=q=0$ for instance and taking $l,j=0,\pm 1$, one discovers
the usual commutation relations of affine Kac Moody ADE Lie algebras
generated by the step operators $\mathrm{T}_{m,0}^{l\alpha }$ ($m\in \mathbb{%
Z})$ and the $\alpha H$ and $K$ commuting Cartan ones. With this choice, the
operator K becomes a central element of the affine algebra while $L$ reduces
to a scaling operator often interpreted as a derivation or again as the zero
mode Virasoso generator $L_{0}$ of 2D conformal algebra. Observe also that
due to the indefinite signature of the bilinear form $\left( x,y\right)
\equiv xy$ of hyperbolic ADE extension, su$\left( 2\right) $ subalgebras
have a remarkable apparent pole singularity.\newline
(c) The Weyl group of hyperbolic extension of affine ADE Lie algebras is
also a semi-direct product generated by reflections and translations and
obeys a quite similar law than the corresponding affine ADE Weyl group. More
remarkable features of eqs(\ref{hypa}) will be given at proper time.

In part II, we consider the geometric engineering of 4D $\mathcal{N}=2$ and $%
\mathcal{N}=1$\ quiver QFTs based on these hyperbolic ADE Lie algebras as
well as their D brane realization. The results obtained here are also used
to study the QFT duals of hyperbolic quiver gauge theories as well as the
analog of RG cascades of affine models.

\section{Indefinite Lie algebras}

By now it established that there exist three principal classes of Lie
algebras; the well known finite dimensional Lie algebras $\mathbf{g}%
_{finite} $, the affine KM algebras $\mathbf{g}_{affine}$ about which we
know quite much and Indefinite Lie algebras $\mathbf{g}_{indef}$ which
continue to hide their secrets mainly because of the large arbitrariness
they contain. Ordinary and affine Lie algebras have sub-classifications
essentially given by the ABCDEFG Cartan classification while there is no
classification yet for indefinite Lie algebras. From a mathematical view,
these three principal classes of Lie algebras are conveniently described in
terms of Cartan matrices $\mathbf{K}_{finite}$, $\mathbf{K}_{affine}$ and $%
\mathbf{K}_{indef} $ respectively. The existence of above three sectors of
Lie algebras is governed by the following central theorem$\cite{m3}$

\subsection{Classification}

We first give the classification theorem and then make a comment.

Theorem:\qquad A generalized indecomposable Cartan matrix $\mathbf{K}$ obeys
\textit{one and only one} of the following three statements: \newline
(\textbf{i}) \textit{Finite type Lie algebras ( }$\det \mathbf{K}>0$ )
characterized by the existence of a real positive definite vector $\mathbf{u}
$ ($u_{i}>0;$ $i=1,2,...$) such that $\mathbf{K}_{ij}u_{j}=v_{j}>0$, where $%
v_{j}$ is a positive vector.\newline
(\textbf{ii}) \textit{Affine type KM algebras, }corank$\left( \mathbf{K}%
\right) =1$, $\det \mathbf{K}=0$\textit{, \ }for which there exist a unique,
up to a multiplicative factor, positive integer definite vector $\mathbf{u}$
( $u_{i}>0;$ $i=1,2,...$) such that $K_{ij}u_{j}=0$. This relation means
that the generalized Cartan matrix $K_{affine}$ has a vanishing eigenvalue.
\newline
(\textbf{iii}) \textit{Indefinite type Lie algebras, ( }$\det K\leq 0$ and
corank$\left( K\right) \neq 1$\textit{, }for which\textit{\ }there exist a
real positive definite vector $\mathbf{u}$ ($u_{i}>0;$ $i=1,2,...$) such
that $K_{ij}u_{j}=-v_{i}<0$, $\ $where $v_{j}$ is as before.

Comment:\qquad In present study we consider a special subset of indefinite
Lie algebras endowed with \textit{symmetrizable} generalized Cartan
matrices. By \textit{symmetrizable, we }mean that corresponding Lie algebras
have a symmetric bi-linear invariant form $\left( ,\right) $ and their
Cartan matrices $K_{ij}$ are realized as,
\begin{equation}
K_{ij}\sim \left( a_{i},a_{j}\right) ,
\end{equation}
where the set $\Pi =\left\{ a_{i}\right\} $ is the set of simple roots to be
discussed later on. For simplicity, we will sometimes refer to the above
product $\left( a_{i},a_{j}\right) $ as $a_{i}a_{j}$.

From above classification theorem, one may already make an idea on
indefinite Lie algebras. For instance, one already feels that roots in
indefinite Lie algebras have much to do with the usual classification of
vectors $\mathbf{v}$ in pseudo-Euclidean spaces $\mathbb{R}^{\left(
p,q\right) }$. There, vectors $\mathbf{v}$ are classified according to their
norms. We have vectors with positive definite norms ($\mathbf{v}^{2}\mathbf{>%
}0$), vectors with zero norms ($\mathbf{\mathbf{v}}^{2}=0$) and vectors
having negative `norms' ($\mathbf{v}^{2}\mathbf{<}0$). As such, one expects
that root systems $\Delta _{indef}$ of indefinite Lie algebras and root
lattices $Q_{indef}$ as well as their maximal toric subalgebras $\mathbf{%
\hbar }_{indef}$ have underlying geometries with indefinite metric $\eta $%
\begin{equation}
\eta =\left(
\begin{array}{cccccc}
-1 &  &  &  &  &  \\
& . &  &  &  &  \\
&  & -1 &  &  &  \\
&  &  & +1 &  &  \\
&  &  &  & . &  \\
&  &  &  &  & +1%
\end{array}
\right) ,
\end{equation}
so that the norm $\mathbf{\mathbf{v}}^{2}=\eta _{ij}v^{i}v^{j}$ which also
reads as $\mathbf{\mathbf{v}}^{2}=-\sum_{i=1}^{p}v_{i}^{2}+%
\sum_{i=p+1}^{q}v_{i}^{2}$ has an indefinite signature, a property which
makes the study of the hyperbolic structure very interesting. The minus sign
in the right hand side of this relation is then an explicit indication of
existence of roots with indefinite norms rendering indefinite Lie algebra
analysis more subtle and more rich. From above Vinberg classification, one
also learns that what we know about Lie algebras is in fact just the top of
an iceberg. For instance what we know on finite dimensional Lie algebras
corresponds just to the deeply Euclidean region of the underlying indefinite
geometry, which naively can be associated with setting $p=0$ in above
relation. In what follows, we focus on the particular hyperbolic subset of
indefinite algebras, its underlying geometry is Lorentzian type ($p=1$).

\subsection{Hyperbolic Lie algebras}

This is a special subset of indefinite Lie algebras which is intimately
related to finite dimensional and affine KM ones. A classification of the
Dynkin diagrams of hyperbolic algebras is available but a long path still
remains to do for the explicit properties of\ the hyperbolic structure. The
results we will give here concerns hyperbolic algebras in the sense of $\cite%
{m5}$; but might be extended to other kinds of indefinite Lie algebras that
go beyond the Wanglai Lie set. Moreover, though closer to ordinary and
affine symmetries, It is interesting to note that the way hyperbolic Lie
algebras enter in quantum physics is still unclear and the role they may
play in the description of QFTs need more explorations. There has been
attempts in this matter during last decade but not enough for a clear
picture. For some specific applications, see $\cite{m51,m52}$ where
hyperbolic algebras are used to characterize a new class of $\mathcal{N}=2$
supersymmetric conformal field theory in four dimensions. For other
applications of hyperbolic algebras, see for instance $\cite{m6}$-$\cite{m13}
$. Now, we turn to give useful details for the present work.

The idea behind the derivation of \textit{hyperbolic} Lie algebras $\mathbf{g%
}_{hyp}$ is based on the same philosophy one uses in building affine Lie
algebras $\mathbf{g}_{affine}$ from finite ones $\mathbf{g}_{finite}$. The
Cartan matrix $K_{hyp}$ of a hyperbolic Lie algebra $\mathbf{g}_{hyp}$ is
obtained in two ways:

(i) Strictly hyperbolic; by starting from the Cartan matrix $K_{finite}$ of
a finite dimensional Lie algebra $\mathbf{g}_{finite}$ and extending it as,
\begin{equation}
K_{finite}\qquad \rightarrow \qquad K_{hyp}=\left(
\begin{array}{cc}
2 & \ast \\
\ast & K_{finite}%
\end{array}
\right)
\end{equation}
where the $\left( \ast \right) $s stand for some row and column vectors.
This kind of extension has no affine KM sub-symmetry; it does not interest
us here and forget about it.

(ii) Hyperbolic, by starting from the Cartan matrix $K_{affine}$ of an
affine KM algebra $\mathbf{g}_{affine}$ and extend it as,

\begin{equation}
\mathbf{K}_{affine}\qquad \rightarrow \qquad \mathbf{K}_{hyp}=\left(
\begin{array}{cc}
2 & \ast \\
\ast & \mathbf{K}_{affine}%
\end{array}
\right) .
\end{equation}
It is these kinds of hyperbolic algebras that we are interested in. Note by
the way that corresponding Dynkin diagrams of $\mathbf{g}_{hyp}$ are
respectively obtained from the underlying ordinary and affine ones by adding
a node $a_{-1}$ as shown on figures1a and 1b.
\begin{figure}[tbh]
\begin{center}
\epsfxsize=7cm \epsffile{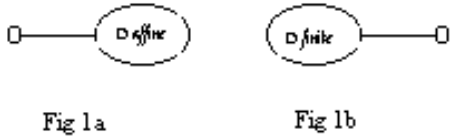}
\end{center}
\caption{{\protect\small \textit{Here we have reported the Dynkin graphs of
the two categories of hyperbolic algebras namely class I (hyperbolic) and
class II (strictly hyperbolic). The box D$_{affine}$\ \ of fig 1a represents
one of the Dynkin diagrams of affine Lie algebras; it is linked to the
hyperbolic node associated with the simple root $a_{-1}$.\ In the second
class, D$_{finite}$\ of fig 1b represents a generic Dynkin diagram of $%
g_{finite}$.}}}
\end{figure}

Following $\cite{m5}$, there are $238$\ possible Dynkin diagrams type those
described by figures 1. These hyperbolic Dynkin diagrams denoted as $%
\mathcal{H}_{i}^{n}$; $i=1,...$, and contain obviously, as sub-diagrams of
co-order $1$, the usual Dynkin graphs associated with $\mathbf{g}_{affine}$
and $\mathbf{g}_{finite}$ Lie algebras.\ By cutting the $a_{-1}$ node of the
order $n$ \textit{hyperbolic} Dynkin diagram, the resulting $\left(
n-1\right) $-\textit{th}\ sub-diagram one gets is either one of the Dynkin
graphs of $\mathbf{g}_{finite}$ or one of $\mathbf{g}_{affine}$ as in above
equation. In what follows we comment the list of the subclass of simply
laced hyperbolic ADE Lie algebras based on affine $ADE$. It is this specific
list of indefinite algebras which concerned here.

\subsection{Hyperbolic ADE Lie algebras}

The full list of simply laced hyperbolic ADE Lie algebras that contain
simply laced affine ADE as a maximal subalgebra is given by,
\begin{eqnarray}
&&\mathcal{H}_{2}^{3},\quad \mathcal{H}_{96}^{3},\quad \mathcal{H}%
_{97}^{3},\quad \mathcal{H}_{98}^{3},\quad \mathcal{H}_{3}^{4},\quad
\mathcal{H}_{1}^{5},\quad \mathcal{H}_{8}^{5},\quad \mathcal{H}%
_{1}^{6},\quad \mathcal{H}_{5}^{6},\quad \mathcal{H}_{6}^{6},  \notag \\
&&\mathcal{H}_{1}^{7},\quad \mathcal{H}_{4}^{7},\quad \mathcal{H}%
_{1}^{8},\quad \mathcal{H}_{4}^{8},\quad \mathcal{H}_{5}^{8},\quad \mathcal{H%
}_{1}^{9},\quad \mathcal{H}_{4}^{9},\quad \mathcal{H}_{5}^{9},\quad \mathcal{%
H}_{1}^{10},\quad \mathcal{H}_{4}^{10}.  \label{na1}
\end{eqnarray}
The Dynkin diagrams of these algebras have remarkable topologies. For the
class of hyperbolic A$_{r}$ algebras, Dynkin diagrams have a loop and look
like the Feynman tade pole diagram of quantum field theory. Hyperbolic DE
algebras have however open topologies involving trivalent vertices.
Sometimes, they are also denoted as T$_{p,q,r}$ or equivalently as $DE_{s}$.
For instance we have $T_{3,2,2}=DE_{5}$. All these Dynkin graphs are simply
laced and obviously of type figure 1a. They turns out to share several
feature with the underlying affine ones. Let us give two illustrating
examples.

\subsubsection{Hyperbolic algebra HA$_{2}$}

This hyperbolic algebra HA$_{2}$ is a leading extension of affine KM algebra
$\widehat{A}_{2}$ which appears as a particular subalgebra. As we will see,
this hyperbolic algebra has four simple roots denoted as $a_{-}$, $a_{0}$, $%
a_{1}$ and $a_{2}$ generating all other roots. To our knowledge, the full
set $\Delta _{hyp}\left( HA_{2}\right) $ of roots of HA$_{2}$ was not worked
out before; it will be given later on. But for the moment, note that $\Delta
_{hyp}\left( HA_{2}\right) $ contains as a proper subset the roots of $%
\widehat{A}_{2}$\ namely,
\begin{eqnarray}
&&\pm \alpha _{1},\qquad \pm \alpha _{2},\qquad \pm \left( \alpha
_{1}+\alpha _{2}\right) ,  \notag \\
&&n\delta ,  \label{ha2} \\
&&n\delta \pm \alpha _{1},\qquad n\delta +\alpha _{2},\qquad n\delta \pm
\left( \alpha _{1}+\alpha _{2}\right)  \notag
\end{eqnarray}
where the first line give the usual roots of ordinary $A_{2}$ and where $%
\delta $ is the familiar imaginary root of affine KM algebras. In eqs(\ref%
{ha2}) $n$ a non zero integer. The $HA_{2}$ algebra is a rank four Lie
algebra and has a $4\times 4$ Cartan matrix given by,

\begin{equation}
\mathbf{K}\left( HA_{2}\right) =\left(
\begin{array}{cccc}
2 & -1 & 0 & 0 \\
-1 & 2 & -1 & -1 \\
0 & -1 & 2 & -1 \\
0 & -1 & -1 & 2%
\end{array}
\right) .
\end{equation}
Its Dynkin diagram, which has four nodes given by the two ordinary ones, the
affine node and the hyperbolic one, is reported on figure 2.
\begin{figure}[tbh]
\begin{center}
\epsfxsize=5cm \epsffile{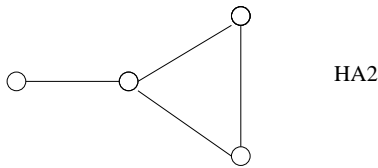}
\end{center}
\caption{{\protect\small \textit{\ This is the Dynkin diagram of}$HA_{2}$.
It has the topology of a Feynman tade pole of QFTs. The node on the left is
associated to the hyperbolic simple root and is linked to the affine one.}}
\end{figure}
From this construction one notes that this hyperbolic algebra has a
trivalent vertex and may be thought of as a kind of gluing together an
affine $\widehat{A}_{2}$ with an ordinary A$_{1}$ at the affine node. The
topology of the graph of HA$_{2}$ shows that such structure is just the
leading term of a more general series involving gluing of affine $\widehat{A}%
_{r}$s with ordinary $A_{m}$s. It can be extended include generalized Dynkin
diagram with generalized Cartan matrix type the following one describing
gluing of affine $\widehat{A}_{2}$ KM algebra with an ordinary $A_{5}$ Lie
algebra.
\begin{equation}
\left(
\begin{array}{cccccccc}
2 & -1 &  &  &  &  &  &  \\
-1 & 2 & -1 &  &  &  &  &  \\
& . & . & . &  &  &  &  \\
&  & . & . & . &  &  &  \\
&  &  & . & 2 & -1 & 0 & 0 \\
&  &  &  & -1 & 2 & -1 & -1 \\
&  &  &  & 0 & -1 & 2 & -1 \\
&  &  &  & 0 & -1 & -1 & 2%
\end{array}
\right)
\end{equation}
Aspects of this kind of symmetries have been considered in $\cite{c3}$. Note
also that the graph of $HA_{2}$ involves a trivalent vertex; which in open
topologies has three linear A$_{i}$ chains as shown on the following
representation
\begin{figure}[tbh]
\begin{center}
\epsfxsize=7cm \epsffile{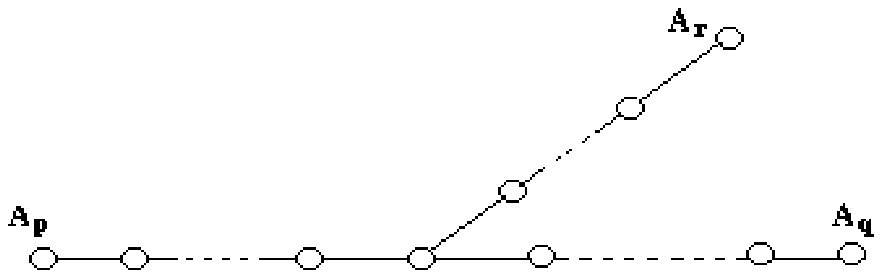}
\end{center}
\caption{{\protect\small \textit{This figure represents a typical vertex in
trivalent mirror geometry. To the central node, it is attached three legs;
two of them are of Dynkin type. The third led is an extra chain which has a
natural interpretation in $T_{p,q,r}$ Lie algebras. These kinds of
topologies are used in the fibration of gauge groups of quiver gauge
theories embedded in Type II on CY3s.}}}
\end{figure}
Recall in passing that graphs involving three ordinary A$_{p}$, A$_{q}$and A$%
_{r}$ Dynkin chains glued at same vertex ( trivalent vertex) are common in
the theory of Lie algebras. For instance the determinant of their
generalized T$_{p,q,r}$ Cartan matrix $K$ reads in general as,
\begin{equation}
\det K\left( T_{p,q,r}\right) =pq+pr+qr-pqr.
\end{equation}
According to the values of p, q and r integers; this determinant can have
all possible signs. These extended Dynkin graph were used few years ago in
the derivation of exact results in $\mathcal{N}=2$ supersymmetric quiver
gauge theories with both fundamental and bi-fundamental matters $\cite{b12}$.

\subsubsection{Hyperbolic algebra HD$_{4}$}

This is a simply laced hyperbolic based on affine $\widehat{D}_{4}$ which
appears as a particular subalgebra. This hyperbolic algebra has also four
simple roots $a_{-}$, $a_{0}$, $a_{1}$ and $a_{2}$ generating all others.
The full set $\Delta _{hyp}\left( HD_{4}\right) $ of roots of H$D_{4}$ will
be given in forthcoming section; it contains as a proper subset the roots of
$\widehat{D}_{4}$\ namely $n\delta +\alpha $ with $n\in \mathbb{Z}$ and $%
\alpha \in \Delta _{finite}\left( D_{4}\right) $. The hyperbolic extension $%
HD_{4}$ has rank five and a $5\times 5$ generalized Cartan matrix given by,

\begin{equation}
K\left( HD_{4}\right) =\left(
\begin{array}{cc}
2 & -1 \\
-1 & K\left( \widehat{D}_{4}\right)%
\end{array}
\right) ,
\end{equation}
where $K\left( \widehat{D}_{4}\right) $\ is the Cartan matrix of affine $%
\widehat{D}_{4}$ and where the $\left( -1\right) $ in first row stands for $%
\left( -1,0,0,0\right) $ and the other one for the transpose vector. Its
Dynkin diagram which is reported on figure 3 has six nodes; four ordinary
ones, one affine and one hyperbolic. \bigskip
\begin{figure}[tbh]
\begin{center}
\epsfxsize=5cm \epsffile{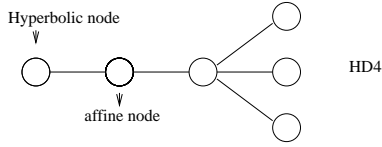}
\end{center}
\caption{{\protect\small \textit{{This is the Dynkin diagram of H}$D_{4}$.
The node on the left corresponding the hyperbolic extension. This diagram
has a tetravalent vertex.}}}
\end{figure}

Here also one can make a similar remark as before; $HD_{4}$ looks as gluing
together three ordinary $A_{1}$s and an ordinary A$_{2}$ at the same vertex.
As far as extension of\ affine ADE Kac Moody algebras are concerned, this
Dynkin diagram can be viewed as just the leading component of a more general
graph involving gluing four A$_{m}$, A$_{p}$, A$_{q}$ and A$_{r}$ chains of
ordinary Lie algebras.
\begin{figure}[tbh]
\begin{center}
\epsfxsize=7cm \epsffile{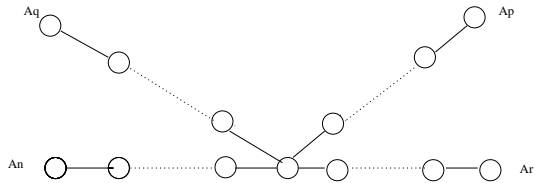}
\end{center}
\caption{{\protect\small \textit{This figure represents a typical vertex in
tetravalent mirror geometry. To the central node, it is attached four $A$
type legs; two of them are of ordinary Dynkin type and the two others are
extra ones.}}}
\end{figure}

\section{Roots in Hyperbolic ADE Lie algebras}

We start by fixing conventional notations we will be using. To avoid
confusion, finite Lie algebras and their affine and hyperbolic extensions
will be denoted respectively as \textbf{g}$_{finite}$, \textbf{g}$_{affine}$
and \textbf{g}$_{hyp}$. Same conventions will be used for corresponding
Cartan matrices $K$, root systems $\Delta $, their simple root basis $\Pi $
and Weyl groups $W$. Note also that \textbf{g}$_{finite}$ stands for one of
the ADE Lie algebras, special properties of these simply laced algebras are
sometimes implicitly used.

\subsection{Triplet Realization}

Let \textbf{g}$_{hyp}$ be a given hyperbolic Lie algebras with an order $n$
Cartan matrix $K_{hyp}$ and a minimal realization involving the following
triplet,
\begin{equation}
\left( \mathbf{\hbar }_{hyp},\Pi _{hyp},\Pi _{hyp}^{v}\right) .  \label{z01}
\end{equation}
As this triplet plays a crucial role in the present construction, let us
comment its contents, fix some useful terminologies and presents the main
lines of our strategy for the reminder of this work.

(\textbf{1})- $\mathbf{\hbar }_{hyp}$ \textbf{space}:\qquad First note that
viewed as a linear set, $\mathbf{\hbar }_{hyp}$ is a complex vector space of
dimension $\left( 2n-m\right) $ with $m=r+2$ being the rank of $K_{hyp}$.
The space $\mathbf{\hbar }_{hyp}$ is endowed with the bilinear form $\left(
,\right) $ introduced in section 2. A generic vector $\hbar $ in $\mathbf{%
\hbar }_{hyp}$ may be then decomposed in terms of some given $e_{i}$ vectors
basis as
\begin{equation}
\hbar =\sum_{i=1}^{2n-m}\hbar _{i}e_{i}.  \label{ex}
\end{equation}
The $\hbar _{i}$s are decomposition coefficients to be interpreted later on
as the commuting Cartan generators of hyperbolic Lie algebras; they may be
defined as usual as,
\begin{equation}
\hbar _{i}=<e_{i}^{\ast },\hbar >,
\end{equation}
where now the $e_{i}^{\ast }$s are the generators of $\mathbf{\hbar }%
_{hyp}^{\ast }$, the dual space of $\mathbf{\hbar }_{hyp}$. Since $\mathbf{%
\hbar }_{hyp}$ and $\mathbf{\hbar }_{hyp}^{\ast }$\ spaces are finite
dimensional, they are isomorphic and so can be identified. The one to one
linear mapping between $\mathbf{\hbar }_{hyp}$ and $\mathbf{\hbar }%
_{hyp}^{\ast }$ is denoted $\upsilon $ and the image of $x\in \mathbf{\hbar }%
_{hyp}$ is $x^{v}\equiv x^{\ast }\in \mathbf{\hbar }_{hyp}^{\ast }$. In the
present case we have $m=n$ and so one can forget about subtleties regarding
the general situations; in particular there is no centre in such hyperbolic
algebras contrary to the usual affine KM algebras where there exist a
central generator K commuting with everything. For further simplicity, we
will focus directly on the physically interesting real subspace,
\begin{equation}
\mathbf{\hbar }_{hyp}=\sum_{i=-1}^{r}\mathbb{R}e_{i},\qquad \mathbf{\hbar }%
_{hyp}^{\ast }=\sum_{i=-1}^{r}\mathbb{R}e_{i}^{\ast },
\end{equation}
with hermitian elements,
\begin{equation}
\hbar ^{\dagger }=\hbar ;\qquad \hbar _{i}^{\dagger }=\hbar _{i}.
\end{equation}
Note that in case $\left\{ e_{i}\right\} $ is an orthogonal basis ($\left(
e_{i},e_{j}\right) =0$ for $i\neq j$), and seen that $\hslash _{hyp}$\ has
to have $\left( 1,r+1\right) $ signature, then the space $\hslash _{hyp}$
formally looks like a $\mathbb{R}^{\left( 1,r+1\right) }$ Lorentzian space
with the metric,
\begin{equation}
\eta _{ij}=diag\left( -1,1,...,1\right) ,
\end{equation}
with $SO\left( 1,r+1\right) $ group as underlying homogeneous symmetry.
Therefore norms $\mathbf{x}^{2}$\ ($\left( \mathbf{x,y}\right) \equiv
\mathbf{xy}$) of vectors $\mathbf{x}$\ in the hyperbolic space $\hslash
_{hyp}$ have an indefinite sign and are as follows
\begin{eqnarray}
e_{-1}^{2} &=&-1,\qquad e_{i}e_{j}=\delta _{ij},\qquad i,j=0,1,...,r,  \notag
\\
\hbar ^{2} &=&\left( -\mathbb{\hbar }_{-1}^{2}+\mathbb{\hbar }%
_{0}^{2}\right) +\sum_{i=1}^{r}\hbar _{i}^{2}.
\end{eqnarray}
It turns out that for hyperbolic Lie algebras based on affine KM symmetries,
it is also convenient to work in light cone frame where $\mathbb{R}^{\left(
1,r+1\right) }$ is thought of as $\mathbb{R}^{\left( 1,1\right) }\oplus
\mathbb{R}^{r}$ with a $SO\left( 1,1\right) \times SO\left( r\right) $
homogeneous symmetry. The previous basis is now changed to $\left\{ e^{\pm
}=\left( e_{0}\pm e_{-1}\right) /\sqrt{2};e_{i},1\leq i\leq r\right\} $
with,
\begin{equation}
\left( e^{\pm }\right) ^{2}=0;\qquad e^{+}e^{-}=1;\qquad e_{i}e_{j}=\delta
_{ij},\qquad i,j=1,...,r.  \label{z1}
\end{equation}
A generic element $\hbar $ of $\mathbf{\hbar }_{hyp}$\ reads in this basis
as $\hbar =K^{+}e^{-}+K^{-}e^{+}+\sum_{i=1}^{r}\hbar _{i}e_{i}$\ and its
norm is
\begin{equation}
\hbar ^{2}=2K^{+}K^{-}+\sum_{i=1}^{r}\hbar _{i}^{2}
\end{equation}
In this way, one recognizes immediately the part $\sum_{i=1}^{r}\hbar
_{i}e_{i}$ of eq(\ref{ex}) as elements of the usual Cartan subspace $\hslash
_{finite}$ and $K^{-}e^{+}$ as the familiar affine central extension of KM
algebras. The extra term $K^{+}e^{-}$ is a new term which has no analogue in
affine Lie algebras; it captures the hyperbolic extension.

(\textbf{2})- \textbf{Root basis}: \qquad As a hyperbolic ADE Lie algebra,
the space $\mathbf{\hbar }_{hyp}$ is generated by root system $\Pi _{hyp}$
rather than the basis (\ref{z1}). Generic elements $\hbar $ are then
expanded with respect to simple root vectors $a_{i}$,

\begin{equation}
\Pi _{hyp}=\left\{ a_{-1},a_{0},a_{1},...,a_{r}\right\} ,  \label{pi}
\end{equation}
with $\left( a_{i},a_{j}\right) =a_{i}a_{j}=K_{ij}$, which is just Cartan
matrix $K_{hyp}$ of the hyperbolic ADE algebras. These $\left( r+2\right) $
simple roots $a_{i}$ are obviously related to previous $e_{i}$s by some
given linear combinations. A convenient choice for our present study
corresponds to take
\begin{equation}
a_{-1}=\gamma -\delta ;\qquad a_{0}=\delta -\psi ,
\end{equation}
while the $r$ other roots $a_{i}$ are as usual in finite dimensional ADE Lie
algebras. Recall for instance that for ordinary A$_{r}$, we have $%
a_{i}=e_{i}-e_{i+1}$. In the above relation, $\psi =\sum_{i=1}^{r}\epsilon
_{i}a_{i}$ is the maximal root of the underlying ordinary ADE Lie algebra
and $\gamma $ and $\delta $ are as follows,
\begin{equation}
\gamma =-e^{-};\qquad \delta =e^{+};\qquad \gamma \delta =-1.
\end{equation}
Notice the extra minus sign used in defining $\gamma $; it is not an adhoc
choice. We will see later that this is the right way to do in order to have
a simpler definition for root positivity in hyperbolic algebras. Note also
that the $\epsilon _{i}$ appearing in $\psi $ are the usual Dynkin weights;
these are positive integers. Turning around the relations $a_{-1}=\gamma
-\delta $ and $a_{0}=\delta -\psi $, it is not difficult to see that $\delta
$ and $\gamma $ are given by the following remarkable sums
\begin{equation}
\gamma =\sum_{i=-1}^{r}\epsilon _{i}a_{i};\qquad \delta
=\sum_{i=0}^{r}\epsilon _{i}a_{i};\qquad \psi =\sum_{i=1}^{r}\epsilon
_{i}a_{i},  \label{ga}
\end{equation}
where for the present study we have $\epsilon _{-1}=\epsilon _{0}=1$. From
this relation, one sees that $\gamma $ is the hyperbolic extension of $%
\delta $ in same manner as $\delta $ is the affine extension of $\psi $. The
above eqs give an idea on how further extension might be done. Moreover as
the hyperbolic simple root $a_{-1}$ should be positive, we clearly see that
positivity of $a_{-1}$ is linked to that of $\gamma $. Positivity of $\gamma
$ follows obviously from positivity of the $\epsilon _{i}$s. Later on, we
will derive a general algorithm for defining root positivity in hyperbolic
algebras.

Using the simple root basis, expansion of generic elements $\hbar $ of
hyperbolic ADE Lie algebras $\mathbf{\hbar }_{hyp}$ reads as,
\begin{equation}
\hbar =K^{+}e^{-}+K^{-}e^{+}+\sum_{i=1}^{r}h_{i}a_{i}=L\gamma +K\delta
+\sum_{i=1}^{r}\mathrm{h}_{i}a_{i}  \label{30}
\end{equation}
where $\left\{ \mathrm{h}_{i};1\leq i\leq r\right\} $ is the usual set of
commuting observables generating the Cartan subalgebra of ordinary ADE and
where $K^{+}$ and $K^{-}$ are two extra hermitian operators dealing with the
affine and hyperbolic extension. Note in passing that though central element
in affine KM subalgebras, the $K^{-}$ generator is no longer a central
element in hyperbolic algebras. The same is valid for $K^{+}$.

(\textbf{3})- \textbf{Coroot basis}:\qquad Note first that, by help of the $%
\upsilon :\mathbf{\hbar }_{hyp}\rightarrow \mathbf{\hbar }_{hyp}^{\ast }$\
isomorphism, the coefficients of the developments eq(\ref{30}), can be
computed by using the dual light cone basis $\left\{ \gamma ^{\ast },\delta
^{\ast },a_{i}^{\ast }\right\} $ and the pairing $<,>$. They are as follows,
\begin{eqnarray}
K^{+} &=&<\gamma ^{\ast },\hbar >=\left( e^{+},\hbar \right) ;\qquad
K^{-}=<\delta ^{\ast },\hbar >=\left( e^{-},\hbar \right) ,  \notag \\
H_{i} &=&<a_{i}^{\ast },\hbar >=K_{ij}h_{j}.
\end{eqnarray}
Like for simple root basis of $\mathbf{\hbar }_{hyp}^{\ast }$ eq(\ref{pi}),
coroot basis $\Pi _{hyp}^{v}$ is a free family of $\mathbf{\hbar }_{hyp}$
defined by,
\begin{equation}
\Pi _{hyp}^{v}=\left\{ a_{-1}^{v},a_{0}^{v},a_{1}^{v},...,a_{r}^{v}\right\} .
\label{z5}
\end{equation}
It allows to express hyperbolic ADE Cartan matrix in term of the pairing
product as $K_{ij}=<a_{i}^{v},a_{j}>$ where
\begin{equation}
a_{i}^{v}=\frac{2}{a_{i}^{2}}\upsilon ^{-1}\left( a_{i}\right)
\end{equation}
Since in ADE Lie algebras $a_{i}^{2}=2$, then $a_{i}^{v}$ can be identified
with $a_{i}$. As such the Cartan matrix reduces to the expression $%
K_{ij}=a_{i}a_{j}$ given before. We end this comment by noting that like in
finite dimensional and affine cases, root system $\Delta _{hyp}$ of
hyperbolic Lie algebras has in general a $\mathbb{Z}_{2}$ gradation implying
that roots can be classified into positive roots and negative ones; i.e
\begin{equation}
\Delta _{hyp}=\Delta _{hyp}^{+}\cup \Delta _{hyp}^{-}  \label{z7}
\end{equation}
Since negative root sub-system $\Delta _{hyp}^{-}$ is just $\left( -\Delta
_{hyp}^{+}\right) $, the main thing one has to deal with is the subset $%
\Delta _{hyp}^{+}$. Moreover, we know from affine KM symmetries that even
within $\Delta _{hyp}^{+}$, one still has to distinguish between two kinds
of roots: (i) real positive roots and (ii) imaginary positive ones. This
implies that $\Delta _{hyp}^{+}$ admits the following sub-grading
\begin{equation}
\Delta _{hyp}^{+}=\Delta _{hre}^{+}\cup \Delta _{him}^{+},  \label{z8}
\end{equation}
where the sub-indices $hre$ and $him$\ refer respectively to hyperbolic real
roots and hyperbolic imaginary ones. Furthermore, we should also have in
mind that $\Delta _{hyp}^{\pm }$, which are spanned by $\Pi _{hyp}$; are in
fact specific subsets of larger space namely the hyperbolic root lattice $%
Q_{hyp}$ and its sub-lattices $Q_{hyp}^{\pm }$,
\begin{equation}
Q_{hyp}^{\pm }=\sum_{i=-1}^{r}\mathbb{Z}^{\pm }a_{i}\subset \mathbf{\hbar }%
_{hyp}.
\end{equation}
The $Q_{hyp}$ lattice\ contains $\Delta _{hyp}$ as a proper subsystem closed
under hyperbolic Weyl transformations to be discussed later on.

\textbf{General Strategy:\qquad }Since by construction, hyperbolic Lie
algebras $\mathbf{g}_{hyp}$ we are interested in here contain as subalgebras
the usual finite dimensional $\mathbf{g}_{finite}$ and affine algebras $%
\mathbf{g}_{affine}$, one may get much information on the structure of
hyperbolic symmetries just by looking for adequate extensions of these
proper sub-symmetries. This will be our strategy not only for (a) Deriving
root system $\Delta _{hyp}$ for hyperbolic ADE algebras which is the main
purpose of this section, but also for: (b) Writing down the explicit form
for the commutation relations of hyperbolic ADE extension, (c) Deriving the
necessary conditions for unitary highest weight representations of these
algebras and (d) Building Weyl symmetries of hyperbolic ADE Lie algebras.
The three last objectives will be described in the forthcoming sections.

\subsection{Building root subsystem $\Delta _{hyp}^{+}$}

From previous discussion, it follows that computation of the full root
contents of $\Delta _{hyp}$ reduces to the determination of $\Delta
_{hre}^{+}$ and $\Delta _{him}^{+}$ sub-systems for real and imaginary
positive roots\footnote{%
The sub-indices $hre$ and $him$ carried by $\Delta _{hre}^{\pm }$ and $%
\Delta _{him}^{\pm }$ \ refer to hyperbolic real and hyperbolic imaginary
respectively. The upper $\pm $ refer to positive and negative roots. Similar
terminology is used for affine root subsystems $\Delta _{are}^{\pm }$ and $%
\Delta _{aim}^{\pm }$. For finite case we have $\Delta _{fre}^{\pm }$ since
there is no $\Delta _{fim}^{\pm }$.}. These are subsets of $Q_{hyp}^{+},$%
\begin{eqnarray}
\Delta _{hre}^{+} &=&\left\{ a=\sum_{i=-1}^{r}k_{i}a_{i}\in
Q_{hyp}^{+}|\qquad \left( a,a\right) >0;\qquad k_{i}\left(
a_{i},a_{i}\right) \in a^{2}\mathbb{Z}\right\} ,  \notag \\
\Delta _{him}^{+} &=&\left\{ a=\sum_{i=-1}^{r}k_{i}a_{i}\in
Q_{hyp}^{+}|\qquad \left( a,a\right) \leq 0\right\} ,  \label{z10}
\end{eqnarray}
which up to now we know only parts of them; that is the part of root $a\in
\Delta _{affine}^{+}$ associated with affine sub-symmetry. What remain to
determine is then the extra part,
\begin{equation}
\Delta _{hyp}^{+}\backslash \Delta _{affine}^{+}.  \label{z11}
\end{equation}
Moreover as $\Delta _{hyp}^{+}$ inherits the Lorentzian signature of $%
Q_{hyp}^{+}$, hyperbolic positive roots are then of three types: space like
roots $a$ with positive definite norms $\left( a,a\right) >0$, positive
light like roots $a$ with zero norms; i.e $\left( a,a\right) =0$ and
positive time like roots $a$ with negative definite `norms' $\left(
a,a\right) <0$\textbf{.} This means that in hyperbolic Lie algebras $\Delta
_{hyp}^{+}$ has a $\mathbb{Z}_{3}$ gradation as,
\begin{equation}
\Delta _{hyp}^{+}=\sum_{q=0,\pm 1}\Delta _{hyp}^{\left( +,q\right) },
\label{z14}
\end{equation}
with,
\begin{eqnarray}
\Delta _{hyp}^{\left( +,+\right) } &=&\left\{ a\in \Delta _{hyp}^{+}|\quad
\left( a,a\right) >0\right\} ,  \notag \\
\Delta _{hyp}^{\left( +,0\right) } &=&\left\{ a\in \Delta _{hyp}^{+}|\quad
\left( a,a\right) =0\right\} ,  \label{z15} \\
\Delta _{hyp}^{\left( +,+\right) } &=&\left\{ a\in \Delta _{hyp}^{+}|\quad
\left( a,a\right) <0\right\}  \notag
\end{eqnarray}
From this gradation, we now know that the usual affine positive roots namely
$\left( 0,n\delta ,\alpha \right) \in \Delta _{are}^{\delta }$ with $n\geq 0$%
, $\alpha \in \Delta _{finite}^{+}$ and $\left( 0,p\delta ,0\right) \in
\Delta _{aim}^{\delta }$ with $p>0$ are distributed in hyperbolic root
system as follows,
\begin{eqnarray}
\left( 0,n\delta ,\alpha \right) &\in &\Delta _{hyp}^{\left( +,+\right) },
\notag \\
\left( 0,p\delta ,0\right) &\in &\Delta _{hyp}^{\left( +,0\right) }.
\label{z16}
\end{eqnarray}
Furthermore, since $\delta $ and $\gamma $ are both of them positive light
like roots and given that they play a perfect symmetric role in the light
cone basis (\ref{z1}), transformations exchanging $\delta $ and $\gamma $
should be a symmetry of $\Delta _{hyp}$. This rotation of $\delta $ and $%
\gamma $ should be an element of the Weyl group as we will later on. This
implies that in $\Delta _{hyp}$ we have not only one affine root sub-system $%
\Delta _{affine}^{\delta }$; but rather two. These are the affine root
sub-systems $\Delta _{affine}^{\delta }$ and $\Delta _{affine}^{\gamma }$
associated with the second light like root $\gamma $. In addition to the
above relations, $\left( n\gamma ,0,\alpha \right) \in \Delta _{are}^{\gamma
}$ with $n\geq 0$, $\alpha \in \Delta _{finite}^{+}$ and $\left( p\gamma
,0,0\right) \in \Delta _{aim}^{\gamma }$ with $p>0$ should be also
hyperbolic positive roots. As such we have the following result,
\begin{eqnarray}
\left( 0,n\delta ,\alpha \right) ;\quad \left( n\gamma ,0,\alpha \right)
&\in &\Delta _{hyp}^{\left( +,+\right) };\qquad \alpha \in \Delta
_{finite}^{+},n\geq 0  \notag \\
\left( 0,p\delta ,0\right) ;\quad \left( p\gamma ,0,0\right) &\in &\Delta
_{hyp}^{\left( +,0\right) };\qquad p>0.  \label{z17}
\end{eqnarray}
These are not all roots that we may have; one has just to note that the
simple root $a_{-1}=\gamma -\delta $ does not figure among these relations.
So there are other roots that still have to be identified. To get these
remaining roots, we will use a remarkable observation and a necessary and
sufficient condition. Let us discuss them separately.

\textbf{Parameterisation}:\qquad\ The observation we refer to above deals
with the fact that roots we know eqs(\ref{z17}) may in general be
parameterized as,
\begin{equation}
a=n\gamma +m\delta +l\alpha ;\qquad \alpha \in \Delta _{finite},  \label{z19}
\end{equation}
with $l=0,\pm 1$ and $m$ and $n$\ are some integers which still need to be
specified. By appropriate choices of these integers, one recovers the
previous results on affine KM systems. Note that as hyperbolic root lattice
should also contain the ordinary one as a proper subset $\ $for any integers
$m$ and $n$; in particular for $m=n=0$,\ it follows that $l\alpha $ must be
a root of ordinary Lie algebra ($l\alpha \in \Delta _{finite}$) which is
possible only if $l=\pm 1$ as usual; but in this case we should also have $%
l=0$. It is worthwhile to note here that this parameterisation is a tricky
one which can be used as well for the derivation of root system of non
simply laced hyperbolic algebras. Recall also that since $\alpha \in \Delta
_{finite}^{+}$ can be expanded in terms of the $r$ simple roots $\alpha
_{i}=a_{i}$, $i=1,...,r$, as $\alpha =\sum_{i=1}^{r}k_{i}a_{i}$ ($k_{i}\in
\mathbb{Z}_{+}$); so eq(\ref{z19}) can be rewritten as,
\begin{equation}
a=n\gamma +m\delta +l\sum_{i=1}^{r}k_{i}a_{i}
\end{equation}
Now replacing $\delta $ and $\gamma $ by their explicit expression (\ref{ga}%
), we get the remarkable expression,
\begin{equation}
a=k_{-1}\gamma +\left( k_{0}-k_{-1}\right) \delta +\sum_{i=1}^{r}\left[
lk_{i}-\left( k_{-1}+k_{0}\right) \right] \epsilon _{i}a_{i}.  \label{rp}
\end{equation}
This formula offers a simple algorithm for defining root positivity in
hyperbolic Lie algebras. The general rule that follows from this algorithm
is: (\textbf{i}) a root type $\left( 0,0,\alpha \right) $ is positive if $%
\alpha $ does; (\textbf{ii}) a root type $a=\left( 0,n\delta ,\alpha \right)
$ is positive if $n$ is positive without reference to $\alpha $ and finally (%
\textbf{iii}) a generic root type $\left( n\gamma ,m\delta ,\alpha \right) $
is positive if the coefficient in front of $\gamma $ does. This feature was
the reason behind the choice of the sign in the identity $\gamma =-e^{-}$.
In term of sings of integers in eq(\ref{rp}), root positivity is expressed
through the condition $k_{-1}>0$. If $k_{-1}=0$; it is manifested through
the condition $k_{0}>0$ and so on. We believe that this algorithm is general
and applies as well for the classification of roots of indefinite Lie
algebras. With these partial results at hand, we turn now to the second point

\textbf{Necessary and sufficient conditions:\qquad }To get the remaining
hyperbolic roots, we start first by recalling a standard lemma on roots of
Lie algebras. Then we use it to identify the remaining roots of $\Delta
_{hyp}$.

\textit{Lemma}:\qquad For a generic hyperbolic root $a$ with an expansion $%
a=\sum_{i=-1}^{r}p_{i}a_{i}$, we have the following results:

\begin{itemize}
\item The family of all real roots $a=\sum_{i=-1}^{r}p_{i}a_{i}$ are such
that $a^{2}>0$ and $p_{i}a_{i}^{2}\in a^{2}\mathbb{Z}$.

\item If a generic real root $a$ verifies $p_{i}a_{i}^{2}\in a^{2}\mathbb{Z}
$ for any $i$; then either $a$ or $\left( -a\right) $ belongs to $\Delta
_{hyp}^{+}$

\item If $a^{2}\leq t$ for some given integer $t$; then either $a$ or $%
\left( -a\right) $ belongs to $\Delta _{hyp}^{+}$.
\end{itemize}

Note that for simply laced hyperbolic Lie algebras where real roots usually
have $a^{2}=a_{i}^{2}=2$, the first and second conditions are trivially
solved. $p_{i}a_{i}^{2}\in a^{2}\mathbb{Z}$ requires that all $p_{i}$s have
to be all of them positive or negative integers; but this is not a new thing
for us since we already know this feature. The novelty comes then from the
third point which translates to our case as,
\begin{equation}
a^{2}\leq 2,
\end{equation}
and turns out to be the necessary and sufficient condition for building the
roots of $\Delta _{hyp}$. Now using our parameterisation $a=n\gamma +m\delta
+l\alpha $, we can express the above necessary and sufficient condition as a
constraint eq on the integer triplet $\left( n,m,l\right) $. This yields,

\begin{equation}
a^{2}=\left( \alpha ^{2}l^{2}-2mn\right) \leq 2.  \label{z20}
\end{equation}
As expected, this constraint eq has a $\mathbb{Z}_{2}\times \mathbb{Z}_{3}$
grading which is carried by the symmetry $a\rightarrow -a$ and the
indefinite sign of the norm $a^{2}$ following from the Lorentzian nature of
the root lattice of hyperbolic algebras.

\textit{Solutions}:\qquad Let us explore the solutions of (\ref{z20}) sector
by sector according to the signs of the norm of roots.

\begin{itemize}
\item \textrm{Space like positive roots}\textbf{:\qquad }In this sector, all
roots are real and have a unique length $a^{2}=2$. Putting this back into eq(%
\ref{z20}), we get the following condition for $\left( n,m,l\right) $,
\begin{equation}
l^{2}-mn=1;\qquad l=0,\pm 1.
\end{equation}
Solutions of this constraint relation are given by the following infinite
set $\left( m,n,l\right) =\left( 0,0,\pm 1\right) ;$ $\left( m,0,\pm
1\right) ;\ \left( 0,n,\pm 1\right) $ and $\left( \mp 1,\pm 1,0\right) $
with $m$ and $n$ non zero integers. Real positive roots of simply laced
hyperbolic ADE algebras read then as
\begin{equation}
m\delta \pm \alpha ;\qquad m\gamma \pm \alpha ;\qquad \gamma -\delta ,
\end{equation}
where $\alpha \in \Delta _{fini}^{+}$ and $m\in \mathbb{Z}_{+}$. The
corresponding negative roots are determined as said before.

\item \textrm{Light like positive roots}\textbf{:\qquad }Hyperbolic light
like roots have a zero norm; $a^{2}=0$. As such the condition eq(\ref{z20})
reduces to $l^{2}-mn=0$ and the solutions for $\left( n,m,l\right) $ yield $%
\left( -1,1,\pm 1\right) ;$ $\left( 1,-1,\pm 1\right) $; $\ \left(
0,m,0\right) \ $and $\left( n,0,0\right) $. Light like positive roots are
then,
\begin{equation}
\gamma +\delta +\alpha ;\qquad m\delta ;\qquad m\gamma ;\qquad \alpha \in
\Delta _{finite}\ ;\quad m\in \mathbb{Z}_{+}^{\ast }
\end{equation}
where now $\alpha \in \Delta _{finite}$\ and $m\in \mathbb{Z}_{+}^{\ast }$

\item \textrm{Time like positive roots}\textbf{:\qquad }Such roots have no
analogue in affine KM algebras because they have negative definite norms;
i.e $a^{2}=-\left\| a^{2}\right\| <0$. Putting back into eq(\ref{z20}), one
gets the condition $2\left( l^{2}-mn\right) =-\left\| a^{2}\right\| $ which
depends on the free parameter $\left\| a^{2}\right\| $ and so has an
arbitrariness which can be used to work out various types of solutions and
so different kinds of hyperbolic extensions. The solutions for $\left(
n,m,l\right) $ will naturally depend on this free parameter and we have a
comment to make here.

\textrm{General solutions}:\qquad\ If all negative lengths are allowed as we
are doing in the present study; i.e $a^{2}<t$ for any integer $t\in \mathbb{Z%
}_{-}$, then the above condition reads as $\left( l^{2}-mn\right) <0$ and
positive time like root solutions are
\begin{eqnarray}
\gamma +m\delta +\alpha ;\qquad m &\in &\mathbb{Z}_{+}-\left\{ 0,1\right\}
\notag \\
n\gamma +m\delta +\alpha ;\qquad m &\in &\mathbb{Z}_{+}^{\ast },\text{ }n\in
\mathbb{Z}_{+}-\left\{ 0,1\right\} \\
n\gamma +m\delta ;\qquad m &\in &\mathbb{Z}_{+}^{\ast },\text{ }n\in \mathbb{%
Z}_{+}^{\ast }.  \notag
\end{eqnarray}
where $\alpha \in \Delta _{finite}$. This is a double infinite set\ showing
that hyperbolic extension involve an extra infinity with respect to the
standard affine one.

\textrm{Particular solutions}:\qquad\ Along with the above general
solutions, there exist also others that are contained in the previous one as
subsets. These correspond to the situations where one replaces the condition
$\left( l^{2}-mn\right) <0$ by weaker ones. This is the case for instance
for the very special example where there is only one negative length $a^{2}$
and this is equal to $\left( -2\right) $. The previous condition
\begin{equation}
2\left( l^{2}-mn\right) =-\left\| a^{2}\right\|
\end{equation}
becomes then $mn=l^{2}+1$ and corresponding time-like positive root
solutions are
\begin{equation}
\gamma +\delta ;\qquad \gamma +2\delta +\alpha ;\qquad 2\gamma +\delta
+\alpha ,\qquad \alpha \in \Delta _{finite}.
\end{equation}
Note that all solutions we have derived are stable under exchanging $\delta $
and $\gamma $. This is not surprising since this property was expected from
the beginning and is manifest in the necessary condition $\left( \alpha
^{2}l^{2}-2mn\right) =a^{2}$ we have used for characterizing hyperbolic
roots. This is then a general feature of the full set $\Delta _{hyp}$; it
reflects just the fact that $\delta $ and $\gamma $ are rotated under Weyl
reflections with respect to $a_{-1}$ as we will see later.
\end{itemize}

\section{More Results}

Consider the $\left( r+2\right) $ dimensional real space $\mathbf{\hbar }%
_{hyp}$ of linear forms introduced in previous section; together with the $%
\mathbf{\hbar }_{affine}$ and $\mathbf{\hbar }_{finite}$ subspaces and the
corresponding lattices $Q_{hyp}$, $Q_{affine}$ and $Q_{finite}$. The
following theorem and corollary give the complete structure of root system
of hyperbolic ADE Lie algebras.

\subsection{Theorem}

Let $\Delta _{finite}$, $\Delta _{affine}$\ and $\Delta _{hyp}$\ be the
sequence of root systems of finite ADE Lie algebras, affine and hyperbolic
extensions satisfying the natural embedding $\Delta _{finite}\subset \Delta
_{affine}\subset \Delta _{hyp}$, then we have the following results:

\begin{itemize}
\item The root system $\Delta _{hyp}$ of hyperbolic ADE Lie algebras belongs
to a Lorentzian lattice; it has a $\mathbb{Z}_{2}\mathbb{\times Z}_{3}$
gradation and so splits into two principal blocs $\Delta _{hyp}^{\pm }$ ($%
\Delta _{hyp}=\Delta _{hyp}^{+}\cup \Delta _{hyp}^{-}$), each one splits in
turn into three subsets as
\begin{equation}
\Delta _{hyp}^{+}=\Delta _{hre}^{\left( +,+\right) }\cup \Delta
_{him}^{\left( +,0\right) }\cup \Delta _{him}^{\left( +,-\right) }
\end{equation}
and
\begin{equation}
\Delta _{hyp}^{-}=\Delta _{hre}^{\left( -,+\right) }\cup \Delta
_{him}^{\left( -,0\right) }\cup \Delta _{him}^{\left( -,-\right) },
\end{equation}
with $\Delta _{hyp}^{-}=\left( -\Delta _{hyp}^{+}\right) $. In these
decompositions, the $\Delta _{hyp}^{\left( +,q\right) }$s with $q=1,0,-1$
are as before; they capture the three regions of the hyperbolic cone.
Sub-indices carried by these $\Delta $s are as before.

\item Root subset $\Delta _{hyp}^{+}$\ is spanned by the positive simple
root system $\Pi _{hyp}=\left\{ a_{i};-1\leq i\leq r\right\} $. In
particular the positive light like roots $\gamma $ and $\delta $, which
belong to $\Delta _{hyp}^{\left( +,0\right) }$,\ are expanded in terms of
simple roots as $\delta =\sum_{i=0}^{r}\epsilon _{i}a_{i}$ and
\begin{equation}
\gamma =\sum_{i=-1}^{r}\epsilon _{i}a_{i},
\end{equation}
where $\epsilon _{i}$ with $i=0,...,r$ are the usual Dynkin numbers of
affine ADE Lie algebras. The extra number $\epsilon _{-1}$ is the weight
associated with $a_{-1}$.

\item The positive light like roots $\gamma $ and $\delta $ are rotated by
the hyperbolic Weyl reflection $\omega _{a_{-1}}\left( x\right) =x-\left(
x,a_{-1}\right) a_{-1}$ as
\begin{equation}
\omega _{-}\left( q\gamma +p\delta \right) =p\gamma +q\delta ,
\end{equation}
where p and q are integers.

\item Root system $\Delta _{hyp}$ contains two isomorphic affine root
sub-systems in one to one correspondence with the two light like roots $%
\delta $ and $\gamma $. These are,
\begin{equation}
\Delta _{affine}^{\delta }=\left\{ m\delta +\alpha ;\quad p\delta ;\quad
\alpha \in \Delta _{finite},\quad m\in \mathbb{Z},\quad p\in \mathbb{Z}%
^{\ast }\right\}  \label{aff1}
\end{equation}
and
\begin{equation}
\Delta _{affine}^{\gamma }==\left\{ n\gamma +\alpha ;\quad q\gamma ;\quad
\alpha \in \Delta _{finite},\quad n\in \mathbb{Z},\quad q\in \mathbb{Z}%
^{\ast }\right\}  \label{aff2}
\end{equation}
which clearly are proper subsets of $\Delta _{hyp}$. As such hyperbolic ADE
algebras have two proper affine ADE subalgebras in one to one with the $%
\delta $ and $\gamma $ pair.

\item Generic hyperbolic roots $a$ in $\Delta _{hyp}$ are generally expanded
in terms of simple roots; but with help of the explicit realization of $%
a_{0} $\ and $a_{-1}$, they may be also represented like,
\begin{equation}
a=k_{-1}\gamma +\left( k_{0}-k_{-1}\right) \delta +\sum_{i=1}^{r}\left(
k_{i}-\left( k_{-1}+k_{0}\right) \epsilon _{i}\right) \alpha _{i}.
\end{equation}
In this representation, root positivity is captured by the positivity of $%
k_{-1}$ whatever the remaining $k_{i}$ integers are. For the special case $%
k_{-1}=0$, this property is transmitted to $k_{0}$ and so on.

\item The root system of simply laced hyperbolic ADE Lie algebras plus the
zero vector ($\Delta _{hyp}\cup \left\{ 0\right\} $) is given by
\begin{equation}
\Delta _{hyp}\cup \left\{ 0\right\} =\left\{ n\gamma +m\delta +l\alpha
\right\} ,  \label{hf}
\end{equation}
where $\alpha \in \Delta _{finite}$where $n$ and $m$ are integers
constrained as $mn\geq \left( l-1\right) \left( l+1\right) $ with integer $l$
taking the three values $l=0,\pm 1$.
\end{itemize}

Now we turn to explore how these relations may be used in the derivation of
the commutation relations.

\subsection{Corollary}

Using the parameterisation eq(\ref{z19}) and given two roots $a=n_{1}\gamma
+m_{1}\delta +l_{1}\alpha $ and $b=n_{2}\gamma +m_{2}\delta +l_{2}\beta $ of
$\Delta _{hyp}$, then we have the following results

\begin{itemize}
\item Generic roots $a=n\gamma +m\delta +l\alpha $ of hyperbolic ADE Lie
algebras obey the necessary and sufficient condition
\begin{equation}
a^{2}\leq 2,
\end{equation}
which reads also as $l^{2}-mn\leq 1$ with $l=0,\pm 1$.

\item The sum $\left( a+b\right) =c$ of two generic hyperbolic roots $a$ and
$b$ is also a hyperbolic root $c=q\gamma +p\delta +k\sigma $ root if and
only if $a^{2}+b^{2}+2ab\leq 2$. Equivalently, we should have
\begin{equation}
k\sigma =l_{1}\alpha +l_{2}\beta ;\qquad p\delta =\left( m_{1}+m_{2}\right)
\delta ;\qquad q\gamma =\left( n_{1}+n_{2}\right) \gamma ,
\end{equation}
with $k=0,\pm 1$ and moreover $k^{2}-pq\leq 1$.

\item The sum $c=a+b$ is: (\textbf{i}) a space like hyperbolic root if $%
a^{2}+b^{2}+2ab=2$ or equivalently $k^{2}-pq=1$; with $k=0,\pm 1$. (\textbf{%
ii}) It is a light like hyperbolic root if $a^{2}+b^{2}+2ab=0$; i.e $%
k^{2}-pq=0$; with $k=0,\pm 1$ and (\textbf{iii}) It is a time like
hyperbolic root if $a^{2}+b^{2}+2ab<0$; i.e $k^{2}-pq<0$; with $k=0,\pm 1$.

\item To each root $a=n\gamma +m\delta +l\alpha $ of $\Delta _{hyp}$, we
associate a step operator $S^{a}$ of the hyperbolic ADE Lie algebras,
\begin{equation}
a=n\gamma +m\delta +l\alpha \qquad \Longleftrightarrow \qquad
S^{a}=S_{m,n}^{l\alpha },  \label{st}
\end{equation}
and to the null vector we associate the Cartan Weyl operator,
\begin{equation}
0=a-a\qquad \Longleftrightarrow \qquad <a,\hbar >=-nL-mK+l\alpha H
\label{st2}
\end{equation}
where $\hbar $ stands for a hermitian element of $\mathbf{\hbar }_{hyp}$.

\item Since to each root $\alpha \in \Delta _{finite}$ is associated a
hyperbolic root $a=n\gamma +m\delta +l\alpha \in \Delta _{hyp}$, it follows
that to the usual invariant number $\alpha ^{2}$ of ordinary ADE Lie
algebras and their affine extension corresponds,.
\begin{equation}
\alpha ^{2}\qquad \Longleftrightarrow \qquad a^{2}=\alpha ^{2}-2mn
\label{nor}
\end{equation}%
With these results at hand we are now ready to build the commutation
relations for hyperbolic ADE Lie algebras.
\end{itemize}

\section{Commutation Relations}

To write down the commutation relations of hyperbolic extension of ADE
algebras, there are at least two ways to follow: an interpolation method and
a covariant approach. Let us first comment these ways to do and then come
back to our main purpose.

\textbf{(1) Interpolation:\qquad }In this explicit way of doing , one thinks
about hyperbolic ADE algebra as an interpolating algebra between its two
affine KM subalgebras so that the generators associated with the imaginary
light like roots $p\delta $ and $q\gamma $\ play a complete symmetric role.
Put differently, the root system $\Delta _{hyp}$ of hyperbolic algebras we
derived before eq(\ref{hf}) is viewed as an interpolating set between its
proper subsets $\Delta _{affine}^{\delta }$\ and $\Delta _{affine}^{\gamma }$%
, eqs(\ref{aff1})-(\ref{aff2}),
\begin{eqnarray}
\Delta _{affine}^{\left( p\delta ,0\right) }\qquad &\Longrightarrow &\qquad
\Delta _{hyp}^{\left( p\delta ,q\gamma \right) }\qquad \Longleftarrow \qquad
\Delta _{affine}^{\left( 0,q\gamma \right) },  \notag \\
&\Uparrow &\qquad \qquad \qquad \qquad \Uparrow \\
\Delta _{affine}^{\left( \delta ,0\right) }\qquad &\Longleftarrow &\qquad
\Delta _{finite}^{\left( 0,0\right) }\qquad \Longrightarrow \qquad \Delta
_{affine}^{\left( 0,\gamma \right) },  \notag
\end{eqnarray}
where $\Delta _{affine}^{\left( \delta ,0\right) }$, $\Delta
_{affine}^{\left( 0,\gamma \right) }$\ and $\Delta _{hyp}^{\left( p\delta
,q\gamma \right) }$\ stand for $\Delta _{affine}^{\delta }$, $\Delta
_{affine}^{\gamma }$\ and $\Delta _{hyp}$ respectively. The set $\Delta
_{finite}^{\left( 0,0\right) }$ is the usual root system $\Delta _{finite}$
of ordinary ADE algebras.

\textbf{(2) Covariant method:\qquad }In this method, one uses the power of
the hyperbolic bilinear form to write down the commutation relations of
hyperbolic algebras in a covariant form. Instead of thinking of roots of
hyperbolic algebras, their norms and corresponding generators as $n\gamma
+m\delta +l\alpha $, $\left( n\gamma +m\delta +l\alpha \right)
^{2}=-2mn+2l^{2}$ and $S_{m,n}^{l\alpha }$, one works directly with vectors $%
a$, norms $a^{2}$ and step operators $S^{\pm a}$ in same manner one does in
finite dimensional Lie algebras. In this way, subalgebras and their root
subsystems correspond to appropriate projections in hyperbolic lattice.

To see how these things work in practice, we start by identifying the
generators of the of hyperbolic ADE algebras and their various su$\left(
2\right) $ subsets.

\subsection{Generators}

As usual, generators of Lie algebras are of two kinds: step operators $%
S^{\pm a}$ and commuting Cartan generators $\frac{2}{a^{2}}a\hbar $ . The
step operators $S^{a}=S_{m,n}^{l\alpha }$ and $S^{-a}=S_{-m,-n}^{-l\alpha }$
eq(\ref{st})\ are associated with roots $\pm a=\pm \left( n\gamma +m\delta
+l\alpha \right) $ belonging to $\Delta _{hyp}$. Their number is same as
order of $\Delta _{hyp}$\ and so there are infinitely many. In quantum
physics, the $S^{-a}$ and $S^{a}$\ operators are interpreted as creation and
annihilation operators and are interchanged under adjoint conjugation
\begin{equation}
\left( S^{\pm a}\right) ^{\dagger }=S^{\mp a};\qquad \left( S_{m,n}^{l\alpha
}\right) ^{\dagger }=S_{-m,-n}^{-l\alpha }\text{.}
\end{equation}
These operators satisfy commutation relations that can be read directly from
the root contents of the hyperbolic system $\Delta _{hyp}$. If one forgets
for a while about poles generated by light like roots, one can already write
down the commutation relations of the usual su$\left( 2\right) $ subalgebras
within $\mathbf{g}_{hyp}$. These are given by,
\begin{equation}
\left[ S^{a},S^{-a}\right] =\frac{2}{a^{2}}a\hbar ;\qquad \left[ \frac{2}{%
a^{2}}a\hbar ,S^{\pm a}\right] =\pm 2S^{\pm a},  \label{su2}
\end{equation}
Before going ahead, note that the pole problem of above relations is not
manifest in Chevalley basis where one is restricted to step operators $%
S^{\pm a_{i}}$ associated with simple roots $\pm a_{i}$. In this
representation, one too simply has,
\begin{equation}
\left[ S^{a_{i}},S^{-a_{i}}\right] =a_{i}\hbar ;\qquad \left[ a_{i}\hbar
,S^{\pm a_{i}}\right] =\pm 2S^{\pm a_{i}}\qquad i=-1,0,1,...,r.
\end{equation}
and it seems that there is no algebraic singularity. This is however not
completely true, the apparent pole difficulty is not really absent but just
translated on the generalized Serre relations defining remaining step
operators. Note also that in the hyperbolic extension we are studying,
commuting Cartan generators are of two types: The $K$, $L$ operators
associated with the light like roots $\delta $ and $\gamma $ and the usual $%
H_{\alpha }$ spaning the Cartan algebra of underlying ordinary ADE
subalgebra. Upon decomposing the root $\alpha $\ on simple root basis ($%
\alpha =\sum_{i=1}^{r}k_{i}\alpha _{i}$), the $H_{\alpha }$s\ can be also
written as $\sum_{i=1}^{r}k_{i}H_{\alpha _{i}}$ or $k_{i}H_{i}$ for
simplicity. These hermitian operators, which satisfy obviously,
\begin{equation}
\left[ K,L\right] =\left[ K,H_{\alpha _{i}}\right] =\left[ H_{\alpha _{i}},L%
\right] =0;\qquad \alpha _{i}\in \Pi _{finite},
\end{equation}
with $H_{\alpha }$\ stands for $\alpha H$, have more than one way to be
handled. They can be handled either separately component by component or
collectively in a compact and covariant form. In the covariant description,
these commuting Cartan generators can put in a useful condensed form as $%
a\hbar $; i.e,
\begin{equation}
a\hbar =-nL-mK+l\alpha H.  \label{cf}
\end{equation}
The point is that since to each pair $\pm a$ of non zero root $\pm a=\pm
\left( n\gamma +m\delta +l\alpha \right) $ belonging to $\Delta _{hyp}$, ($%
a^{2}\leq 2$), one associates an operator triplet $\left\{
S^{a},S^{-a},a\hbar \right\} $, it is then natural that scalar $a\hbar $
which gives the right combination between the $K$, $L$ and $H_{\alpha }$
Cartan generators. This is also dictated by the bilinear form of the
hyperbolic Lie algebras which indicates that the covariantization of $\alpha
H$ should be as in eq(\ref{cf}). Note finally that as this $a\hbar $
operator acts on step operators $S^{\pm b}$ of the hyperbolic algebra as
usual; that is through the adjoint representation,
\begin{equation}
\left[ a\hbar ,S^{\pm b}\right] =\pm abS^{\pm b},
\end{equation}
one sees that on the light cone of root system ($a^{2}=0$), the operator $%
a\hbar $ has zero eigenvalues. This is clearly seen on above relation by
taking $a=b$ and restricting to the light cone of $\Delta _{hyp}$ where $%
a^{2}=0$. But this is exactly what we need to overpass the pole singularity
we have referred to above. This is the reason behind our qualification of
the singularity in
\begin{equation}
\frac{2}{a^{2}}a\hbar =\frac{2}{2mn-l^{2}\alpha ^{2}}\left( nL+mK-l\alpha
H\right) ,  \label{pol}
\end{equation}
is an apparent pole. On light cone of $\Delta _{hyp}$, the eigenvalue of $%
2a^{-2}a\hbar $ behaves as $0/0$; and the indetermination is lifted by
replacing $\frac{2}{a^{2}}a\hbar $ just by $2$. Observe that for ordinary
ADE Lie subalgebras recovered by taking $m=n=0$, the above relation reduces
to the usual one namely,
\begin{equation}
\frac{2}{a^{2}}a\hbar \rightarrow \frac{2}{\alpha ^{2}}\alpha H,
\label{polo}
\end{equation}
and has no pole. The same result is also valid for affine KM subalgebras $%
\mathbf{g}_{affine}^{\delta }$ and $\mathbf{g}_{affine}^{\gamma }$ recovered
by setting $mn=0$ and $m+n\neq 0$. The first corresponds to setting $l=\pm 1$%
, $m=0$ and the second to taking\ $l=\pm 1$, $n=0$. In both cases, eq(\ref%
{pol}) reduces to eq(\ref{polo}). This explains why there is no pole
ambiguity in affine KM algebras. Apparent poles are then a special property
to hyperbolic extensions. Now we are in position to write down the
commutation relations for hyperbolic Lie algebras. We start by the
interpolating method and then we consider the covariant approach.

\subsection{Interpolating Method}

To write down the commutation relations of hyperbolic extension of ADE Lie
algebras, we proceed in three steps as follows: (1) Start by identifying the
usual commutation relations associated with roots in $\Delta _{finite}$;
finite dimensional Lie algebras as subalgebras of the hyperbolic one. (2) We
consider those commutation relations associated with the two special affine $%
\Delta _{affine}^{\delta }$ and $\Delta _{affine}^{\gamma }$ subsets. These
commutation relations describe the two particular isomorphic affine KM ADE
subalgebras $\mathbf{g}_{affine}^{\delta }$ and $\mathbf{g}_{affine}^{\gamma
}$ within hyperbolic ADE generalization. (3) Finally we work out the
commutation relations defining hyperbolic ADE algebras by using
interpolation idea between $\mathbf{g}_{affine}^{\delta }$ and $\mathbf{g}%
_{affine}^{\gamma }$.

\subsubsection{Hyperbolic algebra by gluing pieces}

One of the useful things we have learned from the study of root system $%
\Delta _{hyp}$ is that step operators $S^{\pm a}$ of hyperbolic ADE Lie
algebras carry in general three quantum numbers as
\begin{equation}
S_{m,n}^{\pm l\alpha },
\end{equation}
with $l=0,\pm 1$ and the two other integers $m$ and $n$ are such that $%
l^{2}-nm\leq 1$. For convenience, we split these step operators into two
subsets $\left\{ E_{\pm m,\pm n}^{\pm \alpha }\right\} $ and $\left\{
H_{m,n}^{i}\right\} $ according to whether $l=\pm 1$ or $l=0$. So $E_{\pm
m,\pm n}^{\pm \alpha }$ are the step operators associated with the
hyperbolic roots $\pm a=\pm \left( n\gamma +m\delta +\alpha \right) $ having
an $\alpha $\ dependence ($\pm \alpha \in \Delta _{finite}$) and $%
H_{m,n}^{i} $ are step operators with no explicit $\alpha $\ dependence as
they are associated with roots type $a=n\gamma +m\delta $. The extra upper
index $i=1,...,r$ is related to the rank of simple roots $\alpha _{i}$ in
the underlying finite\ dimensional ADE subalgebra. In addition to the $L$, $%
K $ and $H_{\alpha }=\alpha H$ commuting Cartan observables, the generators
of the hyperbolic ADE algebras we are after are then,
\begin{eqnarray}
E_{m,n}^{\alpha };\qquad &\Leftrightarrow &\qquad a=n\gamma +m\delta +\alpha
,  \notag \\
H_{m,n}^{i};\qquad &\Leftrightarrow &\qquad a=n\gamma +m\delta ,
\end{eqnarray}
where obviously $\alpha \in \Delta _{finite}$ and where the $m$ and $n$
integers are such that $mn\geq l^{2}-1$. We also have the following adjoint
conjugation condition $\left( E_{m,n}^{\alpha }\right) ^{\dagger
}=E_{-m,-n}^{-\alpha }$ and $\left( H_{m,n}^{i}\right) ^{\dagger
}=H_{-m,-n}^{i}$\ useful in the study of the unitary highest weight
representations of hyperbolic ADE algebras. The above set of operators
contain proper subsets that we know quite well as they generate ordinary ADE
subalgebras and their affine extensions. These sets constitute particular
solutions of the constraint eq $mn\geq l^{2}-1$ and are as follows: (i)
finite dimensional subset and two affine KM subalgebras.

\paragraph{Finite dimensional piece}

The finite dimensional piece is generated by the usual zero mode subset of $%
S_{m,n}^{\pm l\alpha }$ corresponding to $l=\pm 1$ and $m=n=0$; that is,
\begin{equation}
\alpha _{i}H\equiv H_{0,0}^{i};\qquad E_{0,0}^{\alpha };\qquad \alpha \in
\Delta _{finite}.
\end{equation}
These operators generate the ordinary ADE Lie subalgebras and their
commutation relations are as follow,
\begin{eqnarray}
\left[ H_{0,0}^{i},H_{0,0}^{j}\right] &=&0,\qquad \left[
H_{0,0}^{i},E_{0,0}^{\alpha }\right] =\alpha ^{i}E_{0,0}^{\alpha },  \notag
\\
\left[ E_{0,0}^{\alpha },E_{0,0}^{\beta }\right] &=&\varepsilon _{\alpha
\beta }E_{0,0}^{\alpha +\beta },\qquad \alpha ,\beta ,\alpha +\beta \in
\Delta _{finite}  \notag \\
\left[ E_{0,0}^{\alpha },E_{0,0}^{-\alpha }\right] &=&\frac{2}{\left( \alpha
,\alpha \right) }\alpha H_{0,0}=\alpha H_{0,0},  \label{ord} \\
\left[ E_{0,0}^{\alpha },E_{0,0}^{\beta }\right] &=&0,\qquad \text{otherwise.%
}  \notag
\end{eqnarray}
where $\varepsilon _{\alpha \beta }$ is the usual antisymmetric tensor.
Observe in passing that the third eq of these relations involves the
bilinear form of finite dimensional Lie algebras $\left( \alpha ,\alpha
\right) $ which cause no problem as it is positive definite ($\alpha ^{2}>0$%
).

\paragraph{Affine pieces}

There are two special isomorphic affine KM subalgebras in our hyperbolic
extension; both of them the extends the previous finite dimensional algebra.
The affine subset corresponding to the case $l=0,\pm 1$ and $n=0$ but $m$ an
arbitrary integer. The corresponding generators are:%
\begin{equation}
L;\qquad mK;\qquad \alpha _{i}H\equiv H_{m,0}^{i};\qquad E_{m,0}^{\alpha
};\qquad \alpha \in \Delta _{finite},\quad m\in \mathbb{Z},
\end{equation}%
and their commutation relations read as follows:
\begin{eqnarray}
\left[ H_{m,0}^{i},H_{p,0}^{j}\right]  &=&-mK\delta ^{ij}\delta _{m+p},
\notag \\
\left[ H_{m,0}^{i},E_{p,0}^{\alpha }\right]  &=&\alpha ^{i}E_{m+p,0}^{\alpha
},  \notag \\
\left[ E_{m,0}^{\alpha },E_{p,0}^{\beta }\right]  &=&\varepsilon _{\alpha
\beta }E_{m+p,0}^{\alpha +\beta },\qquad \alpha ,\beta ,\alpha +\beta \in
\Delta _{finite},  \notag \\
\left[ E_{m,0}^{\alpha },E_{p,0}^{-\alpha }\right]  &=&\frac{2}{\left(
\alpha ,\alpha \right) }\left( \alpha H_{m+p,0}-mK\delta _{m+p}\right) ,
\label{del} \\
\left[ E_{m,0}^{\alpha },E_{p,0}^{\beta }\right]  &=&0,\qquad \text{%
otherwise,}  \notag \\
\left[ K,H_{n,0}^{j}\right]  &=&\left[ K,E_{n,0}^{\alpha }\right] =\left[ K,L%
\right] =0,  \notag \\
\left[ L,H_{m,0}^{i}\right]  &=&mH_{m,0}^{i};\qquad \left[ L,E_{m,0}^{\alpha
}\right] =mE_{m,0}^{\alpha },  \notag
\end{eqnarray}%
Note that here also the fourth eq of these commutation relations involves
the bilinear fom $\left( \alpha ,\alpha \right) =\alpha ^{2}$, which in a
covariant description it should be read as $\alpha ^{2}-2mn$; but this is
exactly $\alpha ^{2}$ since in this case $mn=0$. Along with this affine KM
symmetry, there is also a second proper affine subset which is isomorphic
the above one with roles of integers $m$ and $n$ interchanged. It
corresponds to taking $l=0,\pm 1$ and $m=0$ but $n$ an arbitrary integer. In
this case the generators are,%
\begin{equation}
nL;\qquad K;\qquad \alpha _{i}H\equiv H_{0,n}^{i};\qquad E_{0,n}^{\alpha
};\qquad \alpha \in \Delta _{finite},\quad n\in \mathbb{Z},
\end{equation}%
and the corresponding commutation relations are given by,
\begin{eqnarray}
\left[ H_{0,n}^{i},H_{0,q}^{j}\right]  &=&-nL\delta ^{ij}\delta _{n+q},
\notag \\
\left[ H_{0,n}^{i},E_{0,q}^{\alpha }\right]  &=&\alpha ^{i}E_{0,q+n}^{\alpha
},  \notag \\
\left[ E_{0,n}^{\alpha },E_{0,n}^{\beta }\right]  &=&\varepsilon _{\alpha
\beta }E_{0,q+n}^{\alpha +\beta },\qquad \alpha ,\beta ,\alpha +\beta \in
\Delta _{finite},  \notag \\
\left[ E_{0,n}^{\alpha },E_{0,q}^{-\alpha }\right]  &=&\frac{2}{\left(
\alpha ,\alpha \right) }\left( \alpha H_{0,q+n}-nL\delta _{q+n}\right) ,
\label{gam} \\
\left[ E_{0,n}^{\alpha },E_{0,q}^{\beta }\right]  &=&0,\qquad \text{%
otherwise,}  \notag \\
\left[ L,H_{0,n}^{j}\right]  &=&\left[ L,E_{n,0}^{\alpha }\right] =\left[ L,K%
\right] =0,  \notag \\
\left[ K,H_{0,n}^{i}\right]  &=&nH_{0,n}^{i};\qquad \left[ K,E_{0,n}^{\alpha
}\right] =nE_{0,n}^{\alpha }.  \notag
\end{eqnarray}%
Here also $\alpha ^{2}-2mn$ reduces to $\alpha ^{2}$ since $mn=0$.

\paragraph{Gluing pieces}

From above presentation, one clearly see that eqs(\ref{ord}), (\ref{del})
and (\ref{gam}) appear as special regions of a three dimensional lattice $%
\mathbb{L}\subset $ $\mathbb{Z}^{3}$\ and characterized by the condition $%
mn\geq l^{2}-1$ constraining roots $a$ parameterized like $a=n\gamma
+m\delta +\alpha $\ to be in $\Delta _{hyp}$. We have,
\begin{equation}
\mathbb{L}=\left\{ \left( m,n,l\right) \in \mathbb{Z}^{3}|\quad mn\geq
l^{2}-1;\quad l=0,\pm 1\right\} .  \label{la}
\end{equation}%
In this picture, eqs(\ref{ord}), (\ref{del}) and (\ref{gam}) are
respectively associated with by the sub-lattices $\left\{ \left( 0,0,\pm
1\right) \right\} $, $\left\{ \left( 0,\mathbb{Z},0\right) \cup \left( 0,%
\mathbb{Z},\pm 1\right) \right\} $ and $\left\{ \left( \mathbb{Z},0,0\right)
\cup \left( \mathbb{Z},0,\pm 1\right) \right\} $. From this image, it is not
difficult to wonder what would be the complete set of hyperbolic commutation
relations containing the above ones. Just by trying to glue the pieces eqs(%
\ref{ord}), (\ref{del}) and (\ref{gam}), one may get without major
difficulty the desired result. Indeed, linearity of the algebras requires
that the gluing of the known regions of $\mathbb{L}$\ should be given by
linear interpolating relations between the special subalgebras (\ref{ord})-%
\ref{del}-\ref{gam}). Moreover, covariance of the bilinear form suggests
that whenever there is a scalar product such as $\alpha ^{2}$, it should be
replaced by the covariant one; i.e $\alpha ^{2}-2mn$ instead of $\alpha ^{2}$%
. Therefore hyperbolic extension of affine ADE algebras is roughly speaking
described by the following brackets,%
\begin{eqnarray}
\left[ H_{m,n}^{i},H_{p,q}^{j}\right]  &=&-\left( nL+mK\right) \delta
^{ij}\delta _{m+p}\delta _{n+q},  \notag \\
\left[ H_{m,n}^{i},E_{p,q}^{\alpha }\right]  &=&\alpha
^{i}E_{m+p,n+q}^{\alpha },  \notag \\
\left[ E_{m,n}^{\alpha },E_{p,q}^{\beta }\right]  &=&\varepsilon _{\alpha
\beta }E_{m+p,n+q}^{\alpha +\beta },\qquad \alpha ,\beta ,\alpha +\beta \in
\Delta _{finite},  \notag \\
\left[ E_{m,n}^{\alpha },E_{p,n}^{-\alpha }\right]  &=&\frac{2}{\alpha
^{2}-2mn}\left( \alpha H_{m+p,n+q}-mK\delta _{m+p}\delta _{n+q}-nL\delta
_{m+p}\delta _{n+q}\right) ,  \notag \\
\left[ E_{m,n}^{\alpha },E_{p,q}^{\beta }\right]  &=&0,\qquad \alpha ,\beta
\in \Delta _{finite},\quad \alpha +\beta \notin \Delta _{finite}\text{,}
\notag \\
\left[ K,H_{p,q}^{j}\right]  &=&qH_{p,q}^{j};\qquad \left[ L,H_{p,q}^{j}%
\right] =pH_{p,q}^{j}  \label{inter} \\
\left[ K,E_{p,n}^{\alpha }\right]  &=&nE_{p,n}^{\alpha };\qquad \left[
L,E_{p,q}^{\alpha }\right] =pE_{p,q}^{\alpha },  \notag \\
\left[ K,H_{0,0}\right]  &=&\left[ H_{0,0},L\right] =\left[ K,L\right] =0.
\notag
\end{eqnarray}%
At first sight, these commutation relations seem suffering for light like
roots because of the pole in $\frac{2}{\alpha ^{2}-2mn}$; but this is not a
real difficulty since $\frac{2}{\alpha ^{2}-2mn}$ is not a true singularity
as it can be lifted as discussed before. Having at hand the commutation
relations to give a comment on unitary highest weight representation of this
algebra.

\subsubsection{Necessary conditions for unitary HWRs}

Necessary conditions for unitary highest weight representations of above
hyperbolic algebra may be obtained as usual by looking at the conditions
following from the highest weight representations of its su$\left( 2\right) $
subalgebras,
\begin{equation}
\left[ I_{+},I_{-}\right] =2I_{3};\qquad \left[ I_{3},I_{\pm }\right] =\pm
I_{\pm }
\end{equation}
Since there are as many su$\left( 2\right) $ subalgebras as roots $a=n\gamma
+m\delta +l\alpha $ in the hyperbolic system $\Delta _{hyp}$, we can write
down the corresponding unitary conditions. These are be obtained by
requiring that the eigenvalues of $2I_{3}$, on generic weight vectors $|%
\mathrm{k,l},\mu >$ of the representation space, have to be integral. Here $%
|\mu >$ is a generic weight vector of the underlying ordinary ADE
subalgebras; it is obtained from a highest weight vector $|\lambda >$ by
acting by step operators monomials as,
\begin{equation}
|\mu >=\prod_{\beta \in \Delta _{finie}^{+}}E_{0,0}^{-\beta }|\lambda
>;\qquad E_{0,0}^{\beta }|\lambda >=0;\qquad \alpha H|\lambda >=\alpha
\lambda |\lambda >,
\end{equation}
with $\beta \in \Delta _{finite}$. Moreover as there are two kinds of su$%
\left( 2\right) $ subalgebras in eq(\ref{inter}) according to whether $l=0$
or $l=\pm 1$, it follows that one can write down two types of unitary
necessary conditions. Unitary conditions coming from the block,
\begin{equation}
\left[ H_{m,n}^{i},H_{-m,-n}^{i}\right] =-\left( nL+mK\right) ,  \label{l0}
\end{equation}
interpreted as $\left[ I_{+},I_{-}\right] =2I_{3}$ with no $\alpha H_{0,0}$
term and others coming from
\begin{equation}
\left[ E_{m,n}^{\alpha },E_{-m,-n}^{-\alpha }\right] =\frac{2}{\alpha
^{2}-2mn}\left( \alpha H_{0,0}-mK-nL\right) .  \label{l1}
\end{equation}
In the second case, we have $I_{+}=E_{m,n}^{\alpha }$, $I_{-}=E_{-m,-n}^{-%
\alpha }$ and $2I_{3}=\frac{2}{a^{2}}a\hbar $. Later on, we will see how
both of these relations can be put altogether by help of eq(\ref{cf}); but
for the moment note that part of necessary conditions for unitary highest
weight representations of hyperbolic ADE algebras can be immediately written
down. The point is that as affine KM symmetries are subalgebras of
hyperbolic ADE extension, we should at least have the usual unitary
conditions on the eigenvalues \textrm{k}, \textrm{l} and $\alpha \mu $,
\begin{eqnarray}
\mathrm{k} &=&<\mathrm{k,l,}\mu |K|\mathrm{k,l},\mu >  \notag \\
\mathrm{l} &=&<\mathrm{k,l,}\mu |L|\mathrm{k,l},\mu > \\
\alpha \mu &=&<\mathrm{k,l,}\mu |\alpha H_{0,0}|\mathrm{k,l},\mu >.  \notag
\end{eqnarray}
\ of the operators $K$, $L$ and $\alpha H$ respectively,. Indeed considering
the case where $l=\pm 1$ and setting $m=-1$ and $n=0$ (resp $m=0$ and $n=-1$%
) in eq(\ref{l1}), one sees that the triplet $\left( E_{-1,0}^{\alpha
},E_{1,0}^{-\alpha },2\left( K-\alpha H_{0,0}\right) /\alpha ^{2}\right) $
(resp $\left( E_{0,-1}^{\alpha },E_{0,1}^{-\alpha },2\left( L-\alpha
H_{0,0}\right) /\alpha ^{2}\right) $) form an $su\left( 2\right) $ algebra
and so the eigenvalues of $2\left( K-\alpha H_{0,0}\right) /\alpha ^{2}$
(resp $2\left( L-\alpha H_{0,0}\right) /\alpha ^{2}$) must be integral.
Therefore a first set of necessary conditions reads as,
\begin{eqnarray}
2\left( \mathrm{k}-\alpha \mu \right) &\in &\alpha ^{2}\mathbb{Z};\qquad
\alpha \in \Delta _{finite},  \notag \\
2\left( \mathrm{l}-\alpha \mu \right) &\in &\alpha ^{2}\mathbb{Z};\qquad
\alpha \in \Delta _{finite}
\end{eqnarray}
Moreover, for highest weight states $|\mathrm{k,l},\lambda >$ satisfying
amongst others $E_{1,0}^{-\alpha }|\mathrm{k,l},\lambda >=0$, these
conditions can be reduced further. The idea is that since the commutator $%
\left[ E_{-1,0}^{\alpha },E_{1,0}^{-\alpha }\right] $ on the HW vector $|%
\mathrm{k,l},\lambda >$ is positive because $<\lambda \left[
E_{-1,0}^{\alpha },E_{1,0}^{-\alpha }\right] |\lambda >=\left\|
E_{-1,0}^{\alpha }|\lambda >\right\| ^{2}$, we should also have $\mathrm{k}%
\geq \alpha \lambda \geq 0$ (resp $\mathrm{l}\geq \alpha \lambda \geq 0$).
Therefore, we have the conditions $2\alpha \lambda \in \alpha ^{2}\mathbb{Z}%
_{+}$ and,
\begin{eqnarray}
2\mathrm{k} &\in &\psi ^{2}\mathbb{Z}_{+};\mathbb{\qquad }\mathrm{k}\geq
\psi \lambda ,  \notag \\
2\mathrm{l} &\in &\psi ^{2}\mathbb{Z}_{+};\mathbb{\qquad }\mathrm{l}\geq
\psi \lambda ,
\end{eqnarray}
where $\psi $ is the usual maximal root. Along with these constraint eqs,
there are further constraint eqs coming from the other su$\left( 2\right) $s
within the hyperbolic ADE algebras. We will complete this discussion by
giving the general necessary conditions for unitary highest weight
representations after discussing the covariant approach for hyperbolic
algebras.

\subsection{Covariant method}

Now that we know that the problem of indefinite signature of the bilinear
form is not essential since the pole is just an apparent algebraic
singularity at least at the Lie algebraic level, we can now proceed to write
down the commutation relations for hyperbolic ADE Lie algebras using a
covariant method. Before going into details, it is interesting to note that
despite similarities, hyperbolic ADE algebras differ from what we customary
have in ordinary and affine symmetries. There, roots $a$ have a finite
number of lengths; one length for ordinary ADE and for affine ADE there are
two kinds of root norms; $a^{2}$ is either two or zero. In hyperbolic ADE,
there is an infinite number of possible norms and as we have seen this is
because of the indefinite signature of the bilinear form.

In the covariant method, we will use the generators,
\begin{equation}
S^{a}\equiv S_{m,n}^{l\alpha };\qquad mn\geq l^{2}-1;\qquad l=0,\pm 1,
\end{equation}
instead $H_{m,n}^{i}$ and $E_{p,q}^{\alpha }$ used in the interpolation
approach. We fist give the general form of the commutation relations of our
hyperbolic algebras by using to different but equivalent ways. Then we
complete the discussion on unitary HWRs initiated before.

\subsubsection{Standard basis}

Results from theorem and corollary of section 4 tell us that for any pair of
hyperbolic roots $a=n\gamma +m\delta +l\alpha $ and $b=q\gamma +p\delta
+j\alpha $ belonging to $\Delta _{hyp}$\ system, we have two kinds of\ Lie
algebra generators. The step operators
\begin{eqnarray}
S^{a} &=&S_{m,n}^{l\alpha };\qquad S^{-a}=S_{-m,-n}^{-l\alpha };\qquad
l=0,\pm 1;  \notag \\
S^{b} &=&S_{p,q}^{j\beta },\qquad S^{-b}=S_{-p,-q}^{-j\beta }\qquad j=0,\pm
1,  \label{ss}
\end{eqnarray}
carrying the familiar three quantum numbers $n$, $m$ and $l$ and the $q$, $p$
and $j$ analogue and the usual commuting Cartan generators $L$, $K$ and $H$
which, by help of the bilinear form, can be rewritten in a compact form as,
\begin{eqnarray}
a\hbar &=&-nL-mK+l\alpha H;\qquad l=0,\pm 1,  \notag \\
b\hbar &=&-qL-pK+j\alpha H;\qquad j=0,\pm 1.  \label{ah}
\end{eqnarray}
The commutation relations obeyed by these operators are easily derived; they
follow from the structure of the root system $\Delta _{hyp}$ and are as
follows,
\begin{eqnarray}
\left[ a\hbar ,b\hbar \right] &=&0,\qquad a,\quad b\in \Delta _{hyp},  \notag
\\
\left[ S^{a},S^{b}\right] &=&\varepsilon _{ab}S^{a+b},\qquad a,\quad b,\quad
a+b\in \Delta _{hyp},  \notag \\
\left[ S^{a},S^{-a}\right] &=&\frac{2}{a^{2}}a\hbar ,\qquad a\in \Delta
_{hyp}, \\
\left[ a\hbar ,S^{b}\right] &=&\left( ab\right) S^{b},\qquad a,\quad b\in
\Delta _{hyp},  \notag \\
\left[ S^{a},S^{b}\right] &=&0,\qquad a,\quad b\in \Delta _{hyp},\quad
a+b\notin \Delta _{hyp}.  \notag
\end{eqnarray}
With help of these eqs, one can go ahead an write down the commutation
relations \ in terms of $S_{m,n}^{l\alpha }$ modes. Using the correspondence
eqs(\ref{ah}) and substituting $S^{a}$ and $S^{b}$ by $S_{m,n}^{l\alpha }$
and $S_{p,q}^{j\beta }$ with $\left( l^{2}-mn\right) \leq 1$ and$\left(
j^{2}-pq\right) \leq 1$, we find,
\begin{eqnarray}
\left[ K,H_{0,0}\right] &=&\left[ H_{0,0},L\right] =\left[ K,L\right] =0,
\notag \\
\left[ S_{m,n}^{l\alpha },S_{p,q}^{j\beta }\right] &=&\varepsilon _{l\alpha
,j\beta }S_{m+p,n+q}^{l\alpha +j\beta },\quad \left( l^{2}-mn\right) \leq
1,\left( j^{2}-pq\right) \leq 1,\left( l\alpha +j\beta \right) ^{2}-2\left(
m+p\right) \left( n+q\right) \leq 2,  \notag \\
\left[ S_{m,n}^{l\alpha },S_{-m,-n}^{-la}\right] &=&\frac{2}{\left(
l^{2}\alpha ^{2}-2mn\right) }\left( l\alpha H-nL-mK\right) ,\qquad \left(
l^{2}-mn\right) \leq 1,  \notag \\
\left[ \alpha H,S_{p,q}^{j\beta }\right] &=&j\left( \alpha \beta \right)
S_{p,q}^{j\beta },\qquad \left( j^{2}-pq\right) \leq 1,  \notag \\
\left[ K,S_{p,q}^{j\beta }\right] &=&qS_{p,q}^{j\beta },\qquad \left(
j^{2}-pq\right) \leq 1,  \notag \\
\left[ L,S_{p,q}^{j\beta }\right] &=&pS_{p,q}^{j\beta },\qquad \left(
j^{2}-pq\right) \leq 1,  \label{salg} \\
\left[ S_{m,n}^{l\alpha },S_{p,q}^{j\beta }\right] &=&0,\qquad \left(
l^{2}-mn\right) \leq 1,\quad \left( j^{2}-pq\right) \leq 1,\quad \left(
l\alpha +j\beta \right) ^{2}-2\left( m+p\right) \left( n+q\right) >2,  \notag
\end{eqnarray}
with $l,j=0,\pm 1$ and $\alpha ,\beta \in \Delta _{finite}$. The conditions
on the numbers $n$, $m$ and $l$ and the $q$, $p$ and $j$ ensure that roots $%
a $, $b$ and $a+b$ belong indeed to root system of hyperbolic algebra. By
setting $S_{m,n}^{0}=\frac{1}{r\sqrt{mn}}\sum_{i=1}^{r}H_{m,n}^{i}$ and $%
S_{m,n}^{\alpha }=E_{m,n}^{\alpha }$, it is not difficult to see that above
commutation relations are same as those in eqs(\ref{inter}). Before
preceding further, we would like to make two comments on this hyperbolic
algebra. The first comment concerns unitary conditions for HWRs and the
second deals with link with torus fibration of ordinary ADE.

\paragraph{\textbf{Unitary HWRs}}

Viewed as quantum field theoretical symmetry, the above hyperbolic algebra
seems to have a rich physical spectrum since it has two remarkable branches;
a standard branch and a new one with no analogue in usual affine KM
symmetries. Indeed, unitary highest weight representations of this algebras
require operators $a\hbar $ and $S^{a}$,
\begin{equation}
\left( a\hbar \right) ^{\dagger }=a\hbar ;\qquad \left( S^{a}\right)
^{\dagger }=S^{-a}.  \label{sal1}
\end{equation}
Acting by these operators on weight states $|x>$ ($|x>=|\mathrm{l}\gamma +%
\mathrm{k}\delta +\mu >$) of hyperbolic weight lattice, we have,
\begin{equation}
a\hbar |x>=ax|x>;\qquad S^{-a}|x>=|x+a>,  \label{sal2}
\end{equation}
where the real number $ax$ is expressed in terms of the eigenvalues $\mathrm{%
l}$, $\mathrm{k}$ and $\alpha \mu $ as $ax=\left( -\mathrm{l}m-\mathrm{k}%
n+l\alpha \mu \right) $. Unitary conditions for highest weight
representations of the hyperbolic algebra are obtained as before by
considering the unitary conditions for HW representations of its su$\left(
2\right) $ subalgebras on highest weight states
\begin{equation}
|y>=|\mathrm{l}\gamma +\mathrm{k}\delta +\lambda >.
\end{equation}
Here the state $|\lambda >$ is the same as that we have used before; it
satisfies $E_{0,0}^{\alpha }|\lambda >=0$ and $\alpha H|\lambda >=\left(
\alpha \lambda \right) |\lambda >$ with $\alpha \in \Delta _{finite}$, $%
\frac{2\alpha \lambda }{\alpha \alpha }\in \mathbb{Z}_{+}$. The state $|y>$\
is then just the generalization of ordinary $|\lambda >$ to the of the
hyperbolic weight lattice. It satisfies then,
\begin{equation}
\frac{2}{a^{2}}a\hbar |y>=\frac{2ay}{a^{2}}|y>;\qquad S^{-a}|y>=|y+a>;\qquad
S^{a}|y>=0.  \label{sal3}
\end{equation}
Applying the su$\left( 2\right) $ subalgebra equation $\left[ S^{a},S^{-a}%
\right] =\frac{2}{a^{2}}a\hbar $ on this highest weight state $|y>$, unitary
conditions for HW representations of hyperbolic ADE Lie algebras read in
general as
\begin{equation}
\frac{2ay}{a^{2}}\in \mathbb{Z}_{+}
\end{equation}
for any root $a\in \Delta _{hyp}$. Using the explicit expressions of $a$ and
$y$, the above condition can be also rewritten as follows,
\begin{equation}
\frac{2ay}{a^{2}}=\frac{l\left( \alpha \lambda \right) -\mathrm{l}m-\mathrm{k%
}n}{l^{2}\alpha ^{2}-2mn}\in \mathbb{Z}_{+}\text{ for any }\left(
n,m,l\right) \in \mathbb{L}.  \label{uni}
\end{equation}
Note that for $l=1$ and according to whether $m\neq 0,n=0$ or $m=0,n\neq 0$,
one respectively discovers the usual conditions one has in the case of
unitary highest weight representations of finite dimensional ADE Lie
algebras and their affine extensions. But this is we know and is not a new
thing. Novelty comes rather from the indefinite signature of the bilinear
form leading to two branches since positivity of eq(\ref{uni}) can be solved
in two different ways; either by requiring $ay\geq 0$ and $a^{2}>0$ or $%
ay\leq 0$ and $a^{2}<0$. Put differently, eq(\ref{uni}) is positive if the
two following constraint eqs are fulfilled,
\begin{eqnarray}
l\left( \alpha \lambda \right) -\mathrm{l}m-\mathrm{k}n &\in &\mathbb{Z}%
_{+};\qquad \text{\ and }\qquad l^{2}\alpha ^{2}-2mn\in \mathbb{Z}_{+}^{\ast
},  \notag \\
l\left( \alpha \lambda \right) -\mathrm{l}m-\mathrm{k}n &\in &\mathbb{Z}%
_{-};\qquad \text{\ and }\qquad l^{2}\alpha ^{2}-2mn\in \mathbb{Z}_{-}^{\ast
},
\end{eqnarray}
The first eq is expected as it deals with the deeply Euclidean region ($%
a^{2}>0$) of the hyperbolic root system. The second eq in above relation has
no analogue in affine KM symmetries since ($a^{2}<0$); it captures then the
signature of the hyperbolic structure. Note that along with these two well
defined branches, there is moreover a third special case corresponding to $%
a^{2}=0$. This situation is unclear and deserves more attention.

\paragraph{\textbf{Torus fibration}}

The comment we give here deals with the possible link between hyperbolic
algebra we have been considering and ordinary ADE algebras fibered on two
torus. The point is that our hyperbolic algebra contains two copies of
affine KM symmetries. Each one can be viewed as an ordinary ADE algebras
fibered on a S$^{1}$ cycle. Indeed as we have the habit to do in two
dimensional conformal field theory, the $E_{m,0}^{\pm \alpha }$ and $\alpha
H_{m,0}$ operators ( resp $E_{0,n}^{\pm \alpha }$ and $\alpha H_{0,n}$) have
a nice holomorphic field realization using operator product expansion. There
$E_{m,0}^{\pm \alpha }$ and $\alpha H_{m,0}$ appear as just the Laurent
modes expansion of holomorphic conserved currents $E^{\pm \alpha }\left(
z\right) $ and $\alpha H\left( z\right) $. These are just a fibration of
ordinary ADE generators $E^{\pm \alpha }$ and $\alpha H$ on a S$^{1}$ cycle
as shown below,
\begin{eqnarray}
E^{\pm \alpha }\left( z\right) &=&\sum_{m\in \mathbb{Z}}z^{-m-1}E_{m,0}^{\pm
\alpha };\qquad E_{m,0}^{\pm \alpha }=\oint_{z}\frac{dz}{2i\pi }z^{m}E^{\pm
\alpha }\left( z\right) ,  \notag \\
\alpha H\left( z\right) &=&\sum_{m\in \mathbb{Z}}z^{-m-1}\alpha
H_{m,0};\qquad \alpha H_{m,0}=\oint_{z}\frac{dz}{2i\pi }z^{m}\alpha H\left(
z\right) ,  \label{afcu}
\end{eqnarray}
where, roughly speaking, the complex $z$ variable should be thought of as $%
\left| z\right| =1$. Similar relations may be written down for $E_{0,n}^{\pm
\alpha }$ and $\alpha H_{0,n}$ operators generating the other copy of affine
KM subsymmetry within hyperbolic ADE algebra. The corresponding Laurent
expansion reads as follows,
\begin{eqnarray}
E^{\pm \alpha }\left( w\right) &=&\sum_{n\in \mathbb{Z}}w^{-n-1}E_{0,n}^{\pm
\alpha };\qquad E_{0,n}^{\pm \alpha }=\oint_{w}\frac{dw}{2i\pi }w^{n}E^{\pm
\alpha }\left( w\right) ,  \notag \\
\alpha H\left( w\right) &=&\sum_{n\in \mathbb{Z}}w^{-m-1}\alpha
H_{0,n};\qquad \alpha H_{0,n}=\oint_{w}\frac{dw}{2i\pi }w^{n}\alpha H\left(
w\right) ,
\end{eqnarray}
where now $w$ parameterizes the second S$^{1}$ cycle. Using interpolating
ideas, one may be tempted to think about hyperbolic ADE extension as a
fibration of ordinary ADE generators $E^{\pm \alpha }$ and $\alpha H$ on a
two torus T$^{2}$. Unfortunately this is however not true since if it was
the case the conserved currents would be bi-holomorphic functions $%
S^{l\alpha }\left( z,w\right) $ with Laurent modes
\begin{equation}
S_{m,n}^{l\alpha }=\oint_{z}\frac{dz}{2i\pi }z^{m}\left( \oint_{w}\frac{dw}{%
2i\pi }w^{n}S^{l\alpha }\left( z,w\right) \right) ,  \label{exp}
\end{equation}
defined whatever the $m$ and $n$ integers are. But this is in disagreement
with the constraint eq on hyperbolic roots namely,
\begin{equation}
mn\geq \left( l^{2}-1\right) ;\qquad l=0,\pm 1.
\end{equation}
In the expansion (\ref{exp}), there are more mode operators than allowed by
the above condition. It would be interesting to work out the precise
relation between hyperbolic algebras and torus fibration of finite
dimensional Lie algebras. Seen the difficulty brought by the root constraint
eq $a^{2}\leq 2$ hyperbolic extension of ADE symmetries, we will develop in
what follows a way to overpass, at least, the technical aspect of this
problem.

\subsubsection{New Basis}

Besides indefinite signature of the bilinear form, the second difficulty in
handling hyperbolic algebra comes from the constraint eq $a^{2}\leq 2$ on
allowed roots in $\Delta _{hyp}$. The point is that contrary to affine KM
symmetries, where the solutions for affine constraint eq,
\begin{equation}
a^{2}=\left( l^{2}\alpha ^{2}-2mn\right) =0,2,
\end{equation}
put no limit on $m$ and $n$ integers, the situation is different in
hyperbolic symmetries. General solutions put however constraints on the
allowed values of $m$ and $n$. This Lie algebraic property translates in
terms of 2D CFTs as a constraint eq on the Laurent expansion of
bi-holomorphic functions. A way to insert this behaviour in the hyperbolic
game is to implement the constraint eq $a^{2}=\left( l^{2}\alpha
^{2}-2mn\right) \leq 2$ in the generator basis. Instead of the $S^{a}$
covariant generators considered above, we use rather the new following ones,
\begin{equation}
\mathrm{T}^{a}=Y\left( a^{2}-2\right) S^{a},  \label{ct}
\end{equation}
where $Y\left( x\right) $ is the Heveaside like distribution defined as,
\begin{equation}
Y\left( x\right) =1\text{ \quad if \quad }x\leq 0\text{ \ and \quad }Y\left(
x\right) =0\text{ \quad if \quad }x>0\text{.}
\end{equation}
In this weighted basis $\left\{ \mathrm{T}^{a}\right\} $, the condition $%
a^{2}\leq 2$ is automatically implemented and there is no need to specify at
it a each time. In term of modes, the relation (\ref{ct}), reads as,
\begin{equation}
\mathrm{T}_{m,n}^{l\alpha }=Y\left( l^{2}\alpha ^{2}-2mn-2\right)
S_{m,n}^{l\alpha },  \label{65}
\end{equation}
with $l=0,\pm 1$. Similar relations are also valid for the covariant Cartan
generator $a\hbar $; but for simplicity we will continue to refer to the
normalized operator $Y\left( a^{2}-2\right) a\hbar $ in same manner before.
Now using these new $\mathrm{T}^{a}$ basis operators, one can write down the
covariant expression of the commutation relations of the hyperbolic
extension of the ADE Lie algebras. As before, these relations involve the
indefinite bilinear form $\left( ,\right) $ of the hyperbolic algebra; but
in addition the $Y\left( x\right) $ distribution as shown below,

\begin{eqnarray}
\left[ a\hbar ,b\hbar \right] &=&0,  \notag \\
\left[ a\hbar ,\mathrm{T}^{b}\right] &=&ab\text{ }\mathrm{T}^{b},  \notag \\
\left[ \mathrm{T}^{a},\mathrm{T}^{b}\right] &=&\frac{Y\left( a^{2}-2\right)
Y\left( b^{2}-2\right) }{Y\left( \left( a+b\right) ^{2}-2\right) }%
\varepsilon _{ab}\mathrm{T}^{a+b},  \label{66} \\
\left[ \mathrm{T}^{a},\mathrm{T}^{-a}\right] &=&\frac{2Y\left(
a^{2}-2\right) }{\left( a,a\right) }a\hbar ,  \notag
\end{eqnarray}
Clearly, these commutation relations are linear, antisymmetric, closed and
verify the Jacobi identity. Now replacing the roots by their explicit
expressions; i.e $a=n\gamma +m\delta +l\alpha $, $b=q\gamma +p\delta +j\beta
$ and the sum $a+b$ by $\left( n+q\right) \gamma +\left( m+p\right) \delta
+\left( l\alpha +j\beta \right) H$ and doing the same for the step operators
$\mathrm{T}^{a}$, $\mathrm{T}^{b}$ and $\mathrm{T}^{a+b}$\ which get
replaced by by $\mathrm{T}_{m,n}^{l\alpha }$, $\mathrm{T}_{p,q}^{j\beta }$
and $\mathrm{T}_{m+p,n+q}^{\left( l\alpha +j\beta \right) }$ as in eqs(\ref%
{65}), one can rewrite down the above commutation relations in terms of the
mode operators $\mathrm{T}_{m,n}^{l\alpha }$ and the Cartan generators
namely $nL$, $mK$ and $l\alpha H$. We find,

\begin{eqnarray}
\left[ L,K\right] &=&\left[ K,\alpha H\right] =\left[ \alpha H,\beta H\right]
=0,  \notag \\
\left[ \alpha H,\mathrm{T}_{p,q}^{j\beta }\right] &=&j\left( \alpha \beta
\right) \text{ }\mathrm{T}_{p,q}^{\pm j\beta },  \notag \\
\left[ K,\mathrm{T}_{p,q}^{j\beta }\right] &=&q\mathrm{T}_{p,q}^{j\beta },
\notag \\
\left[ L,\mathrm{T}_{p,q}^{j\beta }\right] &=&p\mathrm{T}_{p,q}^{j\beta },
\label{al} \\
\left[ \mathrm{T}_{m,n}^{l\alpha },\mathrm{T}_{p,q}^{j\beta }\right] &=&%
\frac{Y\left( l^{2}-mn-1\right) Y\left( j^{2}-pq-1\right) \varepsilon
_{l\alpha ,j\beta }}{Y\left( \left( l\alpha +j\beta \right) ^{2}-2\left(
m+p\right) \left( n+q\right) -2\right) }\mathrm{T}_{m+p,n+q}^{\left( l\alpha
+j\beta \right) },  \notag \\
\left[ \mathrm{T}_{m,n}^{l\alpha },\mathrm{T}_{-m,-n}^{-l\alpha }\right] &=&%
\frac{2Y\left( l^{2}-mn-1\right) }{2mn-l^{2}\alpha ^{2}}\left( nL+mK-l\alpha
H\right) ,  \notag
\end{eqnarray}
with $l=0,\pm 1$, $j=0,\pm 1$, $\alpha ,\beta \in \Delta _{finite}$ and
where $l\alpha +j\beta $ should be as $s\eta $ with $s=0,\pm 1$ and $\eta
\in \Delta _{finite}$. From these relations, one recognizes the above
mentioned subalgebras and the unitary conditions for highest weight
representations derived before. In this way of doing, eq(\ref{exp}) extends
as,
\begin{equation}
\mathrm{T}_{m,n}^{l\alpha }=\oint_{z}\frac{dz}{2i\pi }z^{m}\left( \oint_{w}%
\frac{dw}{2i\pi }w^{n}Y\left( l^{2}-mn-1\right) \mathrm{T}^{l\alpha }\left(
z,w\right) \right) .
\end{equation}
where now $\mathrm{T}^{l\alpha }\left( z,w\right) $ are bi-holomorphic
conserved currents. It would be interesting to put this change back in the
algebra (\ref{al}) and try to work out the operator product expansion that
defines the infinite dimensional hyperbolic ADE algebras. This is might be a
way to study field theoretical deformations of two dimensional conformal
field theories with hyperbolic symmetries.

\section{Weyl Groups}

In this section, we want to study the structure of Weyl group $W_{hyp}$ of
the hyperbolic ADE Lie algebras we constructed above. Our interest to this
group comes from recent applications of such kind of structure in the
context of supersymmetric field theories embedded in Type II string
compactification on CY threefolds with ADE singularities. There, the so
called Seiberg like dualities and RG cascades were shown to have a
remarkable interpretation in terms of Weyl transformations. RG cascades,
which do exist in type II strings on CY with affine ADE singularities, were
also shown to be linked with translation symmetries within affine Weyl
groups. We suspect therefore that natural extensions of these Weyl
symmetries to hyperbolic Weyl groups would also have interpretations in the
context of generalized quiver gauge theories such those recently derived in $
\cite{c3,m10}$

\subsection{Strategy towards $W_{hyp}$}

As in affine ADE Lie algebras,\ $W_{hyp}$ groups of hyperbolic ADE Lie
algebras are not defined for all roots $a$ of $\Delta _{hyp}$ just because
generic Weyl reflections $\omega _{a}$ on elements $x$ of the space $\mathbf{%
\hbar }^{\ast }$ are not usually defined. From the following typical Weyl
transformation $\omega _{a}\left( x\right) =x-2\frac{\left( a,x\right) }{%
\left( a,a\right) }a$, we see that this has no sense for light like roots $a$
having zero norm $a^{2}=0$. Thus like in affine case, $W_{hyp}$ is partially
generated by Weyl reflections of finite dimensional Lie algebras and
certainly by translations which still need to be studied. It also has to
leave stable the full hyperbolic root system $\Delta _{hyp}$.

To get the complete structure of the Weyl group of hyperbolic ADE Lie
algebras, we will follow the philosophy we've used in the building of $%
\Delta _{hyp}$; that is by doing things step by step. First by using the
known results on $W_{affine}^{\delta }$ and $W_{affine}^{\gamma }$
respectively associated with the two proper subsets $\Delta
_{affine}^{\delta }$ and $\Delta _{affine}^{\gamma }$ of $\Delta _{hyp}$.
Then by taking advantage of the natural embeddings,
\begin{equation}
W_{hyp}\supset W_{affine}^{\delta }\supset W_{finite},  \label{w1}
\end{equation}
and
\begin{equation}
W_{hyp}\supset W_{affine}^{\gamma }\supset W_{finite},
\end{equation}
as well as symmetries under the inter-change of the imaginary roots $\delta $%
\ and $\gamma $.\ Like in the derivation of root system and the commutation
relations, this way of doing allows us to define $W_{hyp}$ as a parametric
group interpolating between $W_{affine}^{\delta }$ and $W_{affine}^{\gamma }$%
.

\subsection{Interpolation Method}

From Lie algebraic view, Weyl groups are symmetries of root system of
underlying Lie algebras generally acting on vectors $x$ of the dual space $%
\mathbf{\hbar }^{\ast }$ by shifting it as $x+\mu \left( x\right) $. The
shift vector $\mu \left( x\right) $\ is linear in $x$ and its explicit
expression depends on the dimension of the Lie algebra. For the special case
of finite dimensional\ ADE Lie algebras where all roots are space like and
have positive definite norms $a^{2}=2$, Weyl group $W_{finite}$ is a
discrete and finite order symmetry generated by fundamental reflections $%
\omega _{i}$. The latters are associated with simple roots $\alpha _{i}$ and
act on $\mathbf{\hbar }^{\ast }$ as,%
\begin{equation}
\omega _{i}\left( x\right) =x-2\frac{\left( a_{i},x\right) }{\left(
a_{i},a_{i}\right) }a_{i};\qquad i=1,...,r,  \label{s1}
\end{equation}%
which can be further simplified because of the identity $\left(
a_{i},a_{i}\right) =2$. These particular $\omega _{i}$s obey $\omega
_{i}^{2}=I_{id}$ and exhibit special features amongst which we give the
three following:

(\textbf{i}) Non commutative transformations:\qquad\ Though look like
translations, transformations (\ref{s1})\ do not commute since the
composition of two reflections $\omega _{j}$ and $\omega _{i}$ involves
three terms; two of them namely $\left( a_{j},x\right) a_{j}+\left(
a_{i},x\right) a_{i}$ are symmetric as they interchanged under the operation
$a_{i}\leftrightarrow a_{j}$ but the third one $\left( a_{j},a_{i}\right)
\left( a_{i},x\right) a_{j}$ one does not. This property is due to the fact
that the shift $\left( a_{i},x\right) a_{i}$ is non linear in roots; a
property which yields non commutativity of Weyl transformations.

(\textbf{ii}) Group law:\qquad Weyl group of finite Lie algebras has a
finite order; reflections, defined for each node of Dynkin diagram, verify
in general the law $\left( R_{i}R_{j}\right) ^{n_{ij}}=I_{id}$ for some
positive integers $n_{ij}$ depending on the nature of the Lie algebra. It
turns out that these $n_{ij}$s\ are related with the number of link $%
K_{ij}K_{ji}$ from $\alpha _{i}$ to $\alpha _{j}$ nodes of Dynkin diagram.
For a given order $n$ Cartan matrix $K_{ij}$ for instance, we have the
results: $n_{ij}=2$ for the case $K_{ij}K_{ji}=0$, $n_{ij}=3$ for $%
K_{ij}K_{ji}=1$, $n_{ij}=4$ for $K_{ij}K_{ji}=2$ and $n_{ij}=6$ for $%
K_{ij}K_{ji}=3$. These imply, amongst others, that Weyl groups $%
W_{finite}\left( A_{r}\right) $ and $W_{finite}\left( D_{r}\right) $ are
respectively isomorphic to $\mathbb{S}_{r+1}$ and $\mathbb{Z}_{2}^{r}\times
\mathbb{S}_{r};$where $\mathbb{S}_{n}$ stands for permutation group.

(\textbf{iii}) Fix points:\qquad\ A third property, which turns out to be
useful in the derivation of $W_{affine}$ and $W_{hyp}$, is that Weyl
generators $\omega _{i}$s leave the imaginary light like roots $\delta $\
and $\gamma $ as well as their linear combinations $p\delta +q\gamma $
invariants,
\begin{equation}
\omega _{i}\left( p\delta \right) =p\delta ;\qquad \omega _{i}\left( q\gamma
\right) =q\gamma ;\qquad \omega _{i}\left( p\delta +q\gamma \right) =p\delta
+q\gamma ,  \label{w02}
\end{equation}
These kind of roots are obviously absent in finite dimensional Lie algebras;
but appear in affine and hyperbolic root system and turn out to play a basic
role.

Note that because of nilpotency of the norm of $\delta $\ and $\gamma $ ($%
\delta ^{2}=\gamma ^{2}=0$), there is no corresponding Weyl transformation $%
\omega _{\delta }$ nor $\omega _{\gamma }$. Instead of these, there is
rather extra Weyl transformations that have no counterpart in finite
dimensional Lie algebras. These extra symmetries are generated by
translations in the hyperbolic root lattice or more generally in the real
restriction of the space of dual forms $\mathbf{\hbar }^{\ast }$.
Translations distinguish $W_{affine}$\ from $W_{finite}$ and allow to
factories $W_{affine}$ into a semi direct product as follows,
\begin{equation}
W_{affine}^{\delta }=W_{finite}\varpropto T_{\delta }.  \label{w00}
\end{equation}
In this relation $W_{affine}^{\delta }$ stands for affine Weyl group leaving
invariant the affine root subsystem $\Delta _{affine}^{\delta }\cup
\emptyset =\left\{ m\delta +l\beta ;l=0,\pm 1,m\in \mathbb{Z}\right\} $ with
$\beta \in \Delta _{finite}$ and $T_{\delta }=\left\{ t_{\alpha }^{\delta
};\quad \alpha \in \mathbf{\hbar }_{finite}^{\ast }\right\} $ is\ the group
of translations in $\mathbf{\hbar }^{\ast }$ shifting vectors $x$ belonging
to $\mathbf{\hbar }^{\ast }$ as $x+\nu _{\delta }\left( x\right) $ where now
$\nu _{\delta }\left( x\right) $ has a component along the imaginary light
like direction $\delta $. In addition to previous reflections $\omega _{i}$
eq(\ref{s1}), the elements $t_{\alpha }^{\delta }$ generate $W_{affine}$.
Generic elements $t_{\alpha }^{\delta }$ of the set $T_{\delta }$ act on $%
x\in \mathbf{\hbar }_{hyp}^{\ast }$ as,

\begin{equation}
t_{\alpha }^{\delta }\left( x\right) =x+(\delta ,x)\alpha -\left( \alpha
,x\right) \delta -\frac{\alpha ^{2}}{2}\left( \delta ,x\right) \delta ,
\label{e1}
\end{equation}
showing that $\nu _{\delta }\left( x\right) =(\delta ,x)\alpha -\left(
\alpha ,x\right) \delta -\frac{\alpha ^{2}}{2}\left( \delta ,x\right) \delta
$. Before going ahead let us show that these translations are abelian and
exhibit what is the feature that do this job. Acting on the above relation
by $t_{\beta }^{\delta }$ with $\beta \in \Delta _{finite}$ and using eqs(%
\ref{w02}), we get after rearranging terms the following,
\begin{eqnarray}
t_{\beta }^{\delta }\circ t_{\alpha }^{\delta }\left( x\right) &=&x+(\delta
,x)\left[ \beta +\alpha \right] -\left[ \left( \beta +\alpha ,x\right) %
\right] \delta -\frac{\left( \beta +\alpha \right) ^{2}}{2}\left( \delta
,x\right) \delta  \notag \\
&&+(\delta ,x)(\delta ,\beta )\beta -\frac{\beta ^{2}}{2}\left( \delta
,\alpha \right) (\delta ,x)\delta .  \label{w2}
\end{eqnarray}
Moreover as $(\delta ,\beta )=0$ and $\left( \delta ,\alpha \right) =0$
since $\alpha $ and $\beta $\ belong to $\Delta _{finite}$, the second term
of the above equation vanishes identically and consequently $t_{\beta
}^{\delta }\circ t_{\alpha }^{\delta }$ coincides with $t_{\alpha }^{\delta
}\circ t_{\beta }^{\delta }$ which is also $t_{\alpha +\beta }^{\delta }$.
\begin{equation}
t_{\beta }^{\delta }\circ t_{\alpha }^{\delta }\left( x\right) =t_{\alpha
+\beta }^{\delta }=t_{\alpha }^{\delta }\circ t_{\beta }^{\delta }\left(
x\right)
\end{equation}
Commutativity of transformations (\ref{e1}) follows then from the
orthogonality between $\delta $ and $\Delta _{finite}$. This particular
feature gives us the key we need for the derivation of $W_{hyp}$.

\subsection{Hyperbolic Weyl Group $W_{hyp}$}

We start by making three comments which we use for the derivation of the
hyperbolic extension of ADE Weyl groups.

(\textbf{1}) Transformations $t_{\alpha }^{m\delta }\left( x\right) $%
:\qquad\ Translations eq(\ref{e1}) defining the action of $t_{\alpha
}^{\delta }$ on root lattice vectors $x$ may be extended for all imaginary
roots $m\delta $ of the affine root system $\Delta _{affine}^{\delta }$. The
resulting transformations which we denote as $t_{\alpha }^{\left( 0,m\delta
\right) }$ act like,
\begin{equation}
t_{\alpha }^{\left( 0,m\delta \right) }=x+m(\delta x)\alpha -m\left( \alpha
x\right) \delta -\frac{\alpha ^{2}}{2}m^{2}\left( \delta x\right) \delta .
\label{t01}
\end{equation}
Note that this transformation may be also rewritten as $t_{m\alpha }^{\left(
0,\delta \right) }$; but for later use we will keep the first notation.
Composition of two transformations $t_{\alpha }^{\left( 0,m\delta \right) }$
and $t_{\beta }^{\left( 0,p\delta \right) }$ follows in same manner as
before. Straightforward calculation shows that $t_{\beta }^{\left( 0,p\delta
\right) }\circ t_{\alpha }^{\left( 0,m\delta \right) }\left( x\right) $ is
equal to $x+(\delta ,x)\left( p\beta +m\alpha \right) -\left( p\beta
+m\alpha ,x\right) \delta -\frac{\left( p\beta +m\alpha \right) ^{2}}{2}%
\left( \delta ,x\right) \delta ;$ thanks to linearity which implies $\left(
m\delta ,\alpha \right) =m\left( \delta ,\alpha \right) =0$. This
composition is in general equal to $t_{p\beta +m\alpha }^{\left( 0,\delta
\right) }$; but for the special case $m=p$, we also have,
\begin{equation}
t_{\beta }^{\left( 0,m\delta \right) }\circ t_{\alpha }^{\left( 0,m\delta
\right) }=t_{\beta +\alpha }^{\left( 0,m\delta \right) }=t_{\alpha }^{\left(
0,m\delta \right) }\circ t_{\beta }^{\left( 0,m\delta \right) }
\end{equation}
The transformations $t_{\alpha }^{\left( 0,m\delta \right) }$, with $m$
fixed, are obviously abelian for all $\alpha $s belonging to $\Delta
_{finite}$ and more generally in $\mathbf{\hbar }_{finite}$. Note that for $%
m=1$, one recovers the usual defining relation of translation in affine Lie
algebras eq(\ref{s1}); but for $m=0$, the transformation $t_{\alpha
}^{\left( 0,m\delta \right) }$ reduces to the identity operator. This second
feature tells us that there is no analogue of $t_{\beta }^{\left( 0,m\delta
\right) }$ transformations in finite dimensional Lie algebras.

(\textbf{2}) Affine Weyl group $W_{affine}^{\delta }$:\qquad What is valid
for translations involving imaginary roots $m\delta $ \ is also true for $%
n\gamma $ since both of them are orthogonal to $\Delta _{finite}$; $\left(
m\delta ,\alpha \right) =\left( n\gamma ,\alpha \right) =0$. As such one may
also define second kind of translation in hyperbolic space $\mathbf{\hbar }%
_{hyp}^{\ast }$\ as
\begin{equation}
t_{\alpha }^{\left( n\gamma ,0\right) }=x+(n\gamma ,x)\alpha -n\left( \alpha
,x\right) \gamma -\frac{n^{2}\alpha ^{2}}{2}\left( \gamma ,x\right) \gamma .
\label{t02}
\end{equation}
These transformations include the fundamental one $t_{\alpha }^{\left(
\gamma ,0\right) }$ as well as the identity operator $t_{\alpha }^{\left(
0,0\right) }$ of $W_{finite}$. They generate a second abelian subgroup $%
T_{\gamma }$ in analogy with $T_{\delta }$ of eq(\ref{w00}). Naturally, this
deals with the second copy of affine root system $\Delta _{hyp}^{\gamma }$%
contained in $\Delta _{hyp}$ and with a second possible affine Weyl
sub-symmetry $W_{affine}^{\gamma }$ evidently contained in hyperbolic Weyl
group we are looking for. By similarity with eq(\ref{w00}), we have
therefore,

\begin{equation}
W_{affine}^{\gamma }=W_{finite}\varpropto T_{\gamma },  \label{v00}
\end{equation}
with same features as for standard $W_{affine}^{\delta }$.

(\textbf{3}) Hyperbolic Weyl group $W_{hyp}$:\qquad What we have been doing
above is in fact just two aspects of a more general issue. Eqs(\ref{t01})
and (\ref{t02}) are nothing but two special situations of a general picture.
Since any vector $\zeta =m\delta +n\gamma $ is orthogonal to roots $\alpha
\in \Delta _{finite}$\ of finite dimensional Lie algebras, the natural
translations extending eqs(\ref{t01}) and (\ref{t02}) one may define, are
\begin{equation}
t_{\alpha }^{\zeta }=x+(\zeta ,x)\alpha -\left( \alpha ,x\right) \zeta -%
\frac{\alpha ^{2}}{2}\left( \zeta ,x\right) \zeta .
\end{equation}
Using eqs(\ref{w02}) and (\ref{w2}), one can check without difficulty that
these are abelian transformations generating a more general translation set $%
T_{\zeta }=\left\{ t_{\alpha }^{\zeta };\quad \alpha \in \mathbf{\hbar }%
_{finite}^{\ast };\quad \zeta \in \mathbf{\hbar }_{hyp}^{\ast }\backslash
\mathbf{\hbar }_{finite}^{\ast }\right\} $. Previous sets $T_{\delta }$ and $%
T_{\gamma }$ appear as two special situations recovered by taking $\zeta
=\delta $ and $\zeta =\gamma $ respectively. As such we end with the
following structure of the Weyl group $W_{hyp}$ of hyperbolic ADE Lie
algebras,
\begin{equation}
W_{hyp}=W_{finite}\varpropto T_{\zeta }.
\end{equation}
As expected, this group is the semi direct product of $W_{finite}$ with the
co-root lattice $\mathbb{Q}_{hyp}^{v}$ of hyperbolic ADE algebras. It has
quite similar features as $W_{affine}$ and has two subgroups $%
W_{affine}^{\delta }$.and $W_{affine}^{\gamma }$. This construction
generalizes naturally to other hyperbolic extensions of Lie algebras
containing affine symmetries.

\section{Conclusion and Comments}

Motivated by the study of generalizations of affine quiver gauge theories,
we have constructed in this paper the hyperbolic extension of affine ADE
algebras and given necessary conditions of their unitary highest weight
representations. These algebras form a special class of indefinite Lie
algebras and have very remarkable features; some of them can be compared
with their correspondent in affine KM symmetries and many others go beyond.
The present study brings more insight for a better understanding of
solvability of supersymmetric quiver QFTs and their large N duals. The
hyperbolic ADE algebras we have considered here are shown to have no centre,
but have a bi-linear form with indefinite signature making the structure of
these extensions very rich and physically attractive.

Our interest into these hyperbolic generalized Lie algebras came initially
from a tentative to study relevant deformations of supersymmetric quiver
gauge theories emerging as QFT limits in type II compactifications on local
CY manifolds with ADE geometries. But in dealing with this study, we have
noted that except few specific examples, hyperbolic ADE Lie algebras were
surprisingly not enough explored in quantum\ field theoretic literature. For
instance no explicit\ contents of root system of hyperbolic ADE algebras nor
the structure of corresponding Weyl symmetries were used before. Except few
examples, the explicit structure of the commutation relations of these
infinite dimensional algebras as we have the habit to use it in quantum
field theory was also lacking. It was then a necessary task to start by
addressing first these questions before coming to our initial objective.
Among our results in this matter, we mention the five following:\newline
(\textbf{1}) We have derived the explicit contents of root systems $\Delta
_{hyp}$ of hyperbolic ADE symmetries. These are given by the following
double infinite set,
\begin{equation}
\Delta _{hyp}\cup \left\{ 0\right\} =\left\{ n\gamma +m\delta +l\alpha
;\quad mn\geq \left( l^{2}-1\right) ,\quad \alpha \in \Delta _{finite}\quad
l=0,\pm 1\quad m,n\in \mathbb{Z}\right\} ,
\end{equation}%
where $\Delta _{finite}$ stands for the usual root system of finite ADE Lie
algebras and where $\gamma $ and $\delta $ are two light like imaginary
roots satisfying $\left( \gamma ,\delta \right) =-1$, $\left( \gamma ,\gamma
\right) =\left( \delta ,\delta \right) =0$ and $\left( \gamma ,\alpha
\right) =\left( \delta ,\alpha \right) =0$ for $\alpha \in \Delta _{finite}$%
. The known root systems $\Delta _{finite}$ of ordinary ADE symmetries and
their extensions $\Delta _{affine}$ appear naturally as proper subsets as
shown below,
\begin{eqnarray}
\Delta _{hyp} &\supset &\Delta _{affine}^{\delta };\qquad \Delta
_{hyp}\supset \Delta _{affine}^{\gamma },  \notag \\
\Delta _{affine}^{\delta } &\neq &\Delta _{affine}^{\gamma };\qquad \Delta
_{affine}^{\delta }\cap \Delta _{affine}^{\gamma }=\Delta _{finite}.
\end{eqnarray}%
In these embedding, $\Delta _{affine}^{\delta }$ stands for the usual affine
root system $\left\{ \mathbb{Z}\delta +l\alpha ;\quad \alpha \in \Delta
_{finite}\quad l=0,\pm 1\right\} $ and $\Delta _{affine}^{\gamma }$ is an
isomorphic set obtained form $\Delta _{affine}^{\delta }$ by substituting $%
\delta $\ by $\gamma $. It is remarkable that hyperbolic ADE extensions have
two special affine sub-symmetries. The simple root associated with the
affine extension is $\delta -\psi $ for $\Delta _{affine}^{\delta }$ and $%
\gamma -\psi $ for $\Delta _{affine}^{\gamma }$. Imaginary roots of these
two isomorphic affine ADE subalgebras are $m\delta $ and $n\gamma $
respectively. Here $\psi $\ is the maximal root of $\Delta _{finite}$.%
\newline
(\textbf{2}) We have worked out explicitly the defining commutation
relations of hyperbolic ADE Lie algebras by using two basis: (i) a covariant
basis involving manifestly the invariant bi-linear form of the Lorentzian
root lattice Q$_{hyp}$ and (ii) the standard ( non covariant) one, we
usually use in affine ADE KM algebras. We have found, amongst others, that
hyperbolic ADE Lie algebras do indeed have two particular proper affine KM
sub-symmetries $\mathbf{g}_{affine}^{\delta }$ and $\mathbf{g}%
_{affine}^{\gamma }$ with two central extensions K and L,
\begin{eqnarray}
\mathbf{g}_{affine}^{\delta } &\subset &\mathbf{g}_{hyp};\qquad \mathbf{g}%
_{affine}^{\gamma }\subset \mathbf{g}_{hyp},  \notag \\
\mathbf{g}_{affine}^{\delta } &\neq &\mathbf{g}_{affine}^{\gamma };\qquad
\mathbf{g}_{affine}^{\delta }\cap \mathbf{g}_{affine}^{\gamma }=\mathbf{g}%
_{finite},
\end{eqnarray}%
opening as a consequence a window on relevance of hyperbolic ADE extensions
and their possible applications in the study of 2D critical phenomena.
Obviously $\mathbf{g}_{affine}^{\delta }$ and $\mathbf{g}_{affine}^{\gamma }$
are in one to one with the respective root systems $\Delta _{affine}^{\delta
}$ and $\Delta _{affine}^{\gamma }$. From this view, it is also interesting
to note that because of existence of these two remarkable copies of affine
sub-symmetries, hyperbolic ADE Lie algebras $\mathbf{g}_{hyp}$ may be
interpreted as just the interpolating algebra between $\mathbf{g}%
_{affine}^{\delta }$ and $\mathbf{g}_{affine}^{\gamma }$. We suspect that
this idea may have physical interpretations in what one may baptize as
hyperbolic quantum field theoretic systems. It would be interesting to probe
this interpolating idea on the example of supersymmetric quiver gauge
theories following from low energy limit of type II string compactifications
on CY3 with hyperbolic ADE singularities. Axion field of type IIB is
suspected to be behind this interpolation feature. \newline
(\textbf{3}) Turning around results on roots in hyperbolic ADE algebras, we
have found a tricky root particularization ($a=n\gamma +m\delta +l\alpha $)
encoding naturally what we know about ordinary and affine ADE subsystems.
This parallelization allows also to solve the constraint eqs required by
hyperbolic extension in a nice way. In this regards, we have derived the
algorithm for defining positivity of roots in hyperbolic algebras and likely
for indefinite Lie algebras. A generic root $a=n\gamma +m\delta +l\alpha $
of the root system $\Delta _{hyp}$ is said to be positive if $n>0$ whatever
the other $m$ and $l$ integers are. For the special case $n=0$, this
property is transmitted to $m$ which should be positive whatever $l$ integer
is. For $m=0$ the condition is then transmitted to $l$ which again should be
positive. \newline
(\textbf{4}) We have also derived the necessary conditions for unitary
highest weight representations of hyperbolic ADE Lie algebras. Starting from
a generic su$\left( 2\right) $ subalgebra $\left[ S^{a},S^{-a}\right] =\frac{%
2}{a^{2}}a\hbar $ with $a\in \Delta _{hyp}$ and using standard techniques on
unitary highest weight representation theory, we have studied what are the
necessary conditions for unitary representations of hyperbolic ADE algebra.
To do so, we have first shown that from Lie algebraic point of view the pole
$a^{2}=0$ in $\frac{2}{a^{2}}a\hbar $ is not a true singularity as it is
just an apparent difficulty. The point is that the $a\hbar $ observable has
also a zero eigenvalue (ad$a\hbar \left( S^{\pm a}\right) =\pm a^{2}$ $S^{a}$%
) which is exactly what is needed to lift the pole singularity. \ From
representation theory where we have identities type $a\hbar |x>=ax|x>$; see
eqs(\ref{sal1}-\ref{sal3}), the difficulty $\frac{2ax}{aa}$ needs however a
further study before an exact answer. Leaving this point difficulty on
margin, and replacing the generic root $a$ by its explicit expression $%
n\gamma +m\delta +l\alpha $, the above su$\left( 2\right) $ subalgebra
yields,
\begin{equation}
\left[ S_{m,n}^{l\alpha },S_{-m,-n}^{-la}\right] =\frac{2}{\left(
l^{2}\alpha ^{2}-2mn\right) }\left( l\alpha H-nL-mK\right) ,\qquad \left(
l^{2}-mn\right) \leq 1,
\end{equation}%
Acting as usual by these commutation relation on a highest weight vector $%
|y>=|\mathrm{k,l},\lambda >$ of a highest weight representation of
hyperbolic ADE algebra,
\begin{eqnarray}
H|\mathrm{k,l},\lambda &>&=\lambda \text{ }|\mathrm{k,l},\lambda >;\qquad K|%
\mathrm{k,l},\lambda >=\mathrm{k}\text{ }|\mathrm{k,l},\lambda >,  \notag \\
L|\mathrm{k,l},\lambda &>&=\mathrm{l}\text{ }|\mathrm{k,l},\lambda >;\qquad
E_{m,n}^{a}|\mathrm{k,l},\lambda >=0;\qquad n\gamma +m\delta +l\alpha >0,
\end{eqnarray}%
with $H$, $K$ and $L$ being the commuting Cartan generators and $%
S_{m,n}^{l\alpha }$ a generic step operator, one can get unitary necessary
conditions. As such, unitary highest weight representations of hyperbolic
ADE Lie algebras read then as,
\begin{equation}
\frac{l\left( \alpha \lambda \right) -\mathrm{l}m-\mathrm{k}n}{l^{2}\alpha
^{2}-2mn}\in \mathbb{Z}_{+}\text{ for any }\left( n,m,l\right) \in \mathbb{L}%
.  \label{sp}
\end{equation}%
These solutions include as particular cases $\mathrm{k}\geq \psi \lambda $
and $\mathrm{l}\geq \psi \lambda $ which one recognizes as the unitary
conditions for highest weight representations of the two proper affine
sub-symmetries. The above conditions follow naturally by using the same su$%
\left( 2\right) $ trick one uses in the derivation of unitary
representations of finite dimensional and affine Lie algebras.\newline
(\textbf{5}) Finally we have constructed the Weyl group of the hyperbolic
extension of affine ADE Lie algebras. This group which is shown to be given
by,
\begin{equation}
W_{hyp}=W_{finite}\varpropto T_{\zeta },
\end{equation}%
is the semi direct product of $W_{finite}$ with the co-root lattice $\mathbb{%
Q}_{hyp}^{v}$ of hyperbolic ADE algebras. In addition to the Weyl group $%
W_{finite}$ of finite ADE Lie algebras, $W_{hyp}$ has as expected two
special isomorphic proper subgroups $W_{affine}^{\delta }$.and $%
W_{affine}^{\gamma }$.
\begin{eqnarray}
W_{affine}^{\delta } &\subset &W_{hyp};\qquad W_{affine}^{\gamma }\subset
W_{hyp},  \notag \\
W_{affine}^{\delta } &\neq &W_{affine}^{\gamma };\qquad W_{affine}^{\delta
}\cap W_{affine}^{\gamma }=W_{finite}.
\end{eqnarray}%
It is these groups which are used in $\cite{brs}$; see also $\cite{fiol}$,
to study Seiberg like dualities and RG cascades in hyperbolic quiver gauge
theories.

Along with the result we have obtained in this paper, we have also a comment
to add. Besides the obvious fact that the present work generalizes naturally
to other hyperbolic extensions of Lie algebras; in particular to hyperbolic
algebras based on non simply laced affine symmetries, we have learned an
important lesson which can serve as a guide for dealing with\ Indefinite Lie
algebras and their quantum field theoretical realizations. Through the above
analysis we have learnt that complexity due to indefinite signature of the
bilinear form is only apparent. Much things on the study of hyperbolic
extension of ordinary Lie algebras can be done using a similar philosophy as
in special relativity and quantum electrodynamics QED on Minkowski space. In
particular roots are of three kinds; space like roots with positive definite
norms, light like and time like ones with negative norms. It is then not
surprising if structure constants of hyperbolic ADE Lie algebras and their
representations capture the above cone details. As far as quantum physics
realizations and unitary highest weight representations of hyperbolic
algebras are concerned, it is worthwhile to mention that the spectrum
following from eq(\ref{sp}) is no so strange as one may think. The unique
novelty with respect to what we know about ordinary and affine Lie algebras
is that now the spectrum is richer and so more interesting. For instance the
unitary conditions eq(\ref{uni})-(\ref{sp}),
\begin{equation}
2\frac{ay}{aa}\in \mathbb{Z}_{+}.
\end{equation}%
involves now two branches instead of one. Since the bilinear form is
indefinite, positivity of this relation implies: (i) an ordinary class of
solutions where both $ay$ and $aa$ are positive; it contains as particular
subsets what we know on finite dimensional ADE Lie algebras and their affine
extensions. (ii) a new class of solutions where both of $ay$ and $aa$ are
negative so that their ratio is a positive integer. This second branch has
no analog in ordinary and affine Lie algebras.

\begin{acknowledgement}
The authors would like to thank Dr Adil Belhaj for earlier collaboration on
this matter. Saidi thanks Prof Cesar Gomez for kind hospitality at IFT/CSIC,
Madrid. This work enters in the frame work of D12/25, Protars III CNRT (
Rabat).
\end{acknowledgement}

\end{document}